\documentclass[12pt]{article}

\usepackage{scicite,color}
\usepackage[normalem]{ulem}

\usepackage{times}
\usepackage{graphicx,amsmath,amssymb}

\usepackage{verbatim}

\usepackage{times, bbold, bm}

\newcommand{\bk}{{\bm k}}
\newcommand{\cH}{{\cal H}}
\newcommand{\cF}{{\cal F}}

\newenvironment{sciabstract}{%
\begin{quote} \bf}
{\end{quote}}

\title{\vspace{-60pt}Emergent Interfacial Superconductivity between Twisted Cuprate Superconductors} 

\author
    {S. Y. Frank Zhao$^1$, Nicola Poccia$^{2,1}$, Xiaomeng Cui$^1$, Pavel A. Volkov$^3$,\\Hyobin Yoo$^{1}$\footnote{Current Address: Department of Physics, Sogang University, Seoul, Korea}, Rebecca Engelke$^1$, Yuval Ronen$^1$, Ruidan Zhong$^5$ \footnote{Current Address: Tsung-Dao Lee Institute and School of Physics and Astronomy, Shanghai Jiao Tong University, Shanghai 200240, China}, \\Genda Gu$^5$, Stephan Plugge$^6$, Tarun Tummuru$^6$, \\Marcel Franz$^6$, Jedediah H. Pixley$^3$, Philip Kim$^{1\S}$
    \\
    \normalsize{$^{1}$Department of Physics, Harvard University, Cambridge, MA 02138, USA}\\
    \normalsize{$^{2}$Institute for Metallic Materials, IFW Dresden, 01069 Dresden, Germany}\\
    \normalsize{$^{3}$Department of Physics and Astronomy, Center for Materials Theory,} \\ \normalsize{Rutgers University, Piscataway, NJ 08854, USA}\\
    \normalsize{$^{5}$Department of Condensed Matter Physics and Materials Science,}\\  
    \normalsize{Brookhaven National Laboratory, Upton, NY 11973, USA}\\
    \normalsize{$^{6}$Department of Physics and Astronomy \& Stewart Blusson Quantum Matter Institute,}\\ 
    \normalsize{University of British Columbia, Vancouver, BC, Canada}\\
    \normalsize{$^{\S}$To whom correspondence should be addressed; E-mail: pkim@physics.harvard.edu}
}

\date{}

\begin{document} 
\baselineskip20pt

\maketitle 



\begin{sciabstract} 
   Twisted interfaces between stacked van der Waals cuprate crystals enable tunable Josephson coupling between in-plane anisotropic superconducting order parameters. Employing a novel cryogenic assembly technique, we fabricate Josephson junctions with an atomically sharp twisted interface between Bi$_2$Sr$_2$CaCu$_2$O$_{8+x}$ crystals. 
 The Josephson critical current density sensitively depends on the twist angle, reaching the maximum value comparable to that of the 
    intrinsic junctions at small twisting angles, and is suppressed by almost 2 orders of magnitude yet remains finite close to 45$^\circ$ twist angle. 
    Through the observation of fractional Shapiro steps and the analysis of Fraunhofer patterns we show that the remaining superconducting coherence near $45^\circ$ is due to the co-tunneling of Cooper pairs, a necessary ingredient for high-temperature topological superconductivity.

\end{sciabstract}

Weak van der Waals (vdW) bonding between neighboring atomic layers offers a unique opportunity for engineering atomic interfaces with controlled twist angles~\cite{ribeiro-palau2018}. Careful adjustment of the twist angle can create the spatial periodicity of a moir\'e superlattice~\cite{yoo2019} with narrow electronic bands and topological structure~\cite{carr2020}. Realizations of such \textit{‘twistronics’} host a plethora of emergent electronic states, including superconductivity \cite{cao2018}, magnetism \cite{lu2019}, Chern insulators \cite{saito2021}, generalized electronic Wigner crystals \cite{zhou2021}, and correlated insulating states \cite{cao2018correlated} at the twisted interface of various vdW materials, including graphene~\cite{carr2020}  and transition metal dichalcogenides \cite{kennes2021}. 

Atomically layered cuprate high temperature superconductors also offer a platform for twistronics by engineering the coupling between nodal superconducting order parameters (SOP) across a twisted vdW interface~\cite{sigrist1998, GeneralTwist, canHightemperatureTopologicalSuperconductivity2021, Volkov, TwistTheoryRev}. In Bi$_2$Sr$_2$CaCu$_2$O$_{8+x}$ (BSCCO), superconducting CuO$_2$ bilayers are Josephson-coupled through insulating [SrO-BiO] bilayers~\cite{IntrinsicStructure}, where the crystal can be mechanically cleaved into atomically flat crystals \cite{BSCCOHall, BSCCONanoLett} exhibiting high temperature superconductivity even in the monolayer limit~\cite{MonolayerBSCCO}.

Twisted interfacial Josephson junctions (JJ) between superconductors directly probe the pairing symmetry of Cooper pairs. In principle, interfacial Josephson coupling between twisted nodal $d$-wave superconductors is strongly modulated by the twist angle~\cite{TwistTheoryRev}. At exactly $\theta = 45^\circ$, direct Cooper pair tunneling is forbidden due to the complete mismatch between the $d_{x^2-y^2}$ symmetric SOPs across the interface~\cite{TwistTheoryRev}. The second-order co-tunneling of Cooper pairs is allowed \cite{Goldobin, sigrist1998, GeneralTwist, canHightemperatureTopologicalSuperconductivity2021}, and is expected to support topological, time-reversal symmetry (TRS) breaking superconducting phases persisting up to the junction superconducting transition temperature $T_C$~\cite{canHightemperatureTopologicalSuperconductivity2021, GeneralTwist}. Alternatively, TRS can be broken away from $\theta = 45^\circ$ via an applied current, which also induces a topological superconducting state~\cite{Volkov}.
	
The preservation of surface superconductivity of BSCCO crystals after vdW stacking remains an outstanding experimental challenge \cite{BSCCOHall, BSCCONanoLett}. BSCCO crystals react with moisture \cite{MonolayerBSCCO, sandilands2014} and their oxygen dopants become mobile above $200$ K \cite{fratiniScalefreeStructuralOrganization2010, MonolayerBSCCO}. BSCCO twist junctions required high temperature oxygen annealing to restore interfacial superconductivity \cite{TwistJctnBulk, TwistJctnWhisker, TwistJctnFlakes, TwistJctnDWave}, often at the cost of significant interfacial structural reconstruction \cite{TwistJctnBulkTEM,TwistJctnFlakes}. The majority of experiments observed no angular sensitivity \cite{TwistJctnBulk,TwistJctnWhisker,TwistJctnFlakes}, except one \cite{TwistJctnDWave} where the Josephson coupling angular dependence deviated strongly from conventional models of $d$-wave superconductivity.

\begin{figure*}[!ht]
	\vspace{-0.5cm}
	\begin{center}
		\includegraphics[width=1\linewidth]{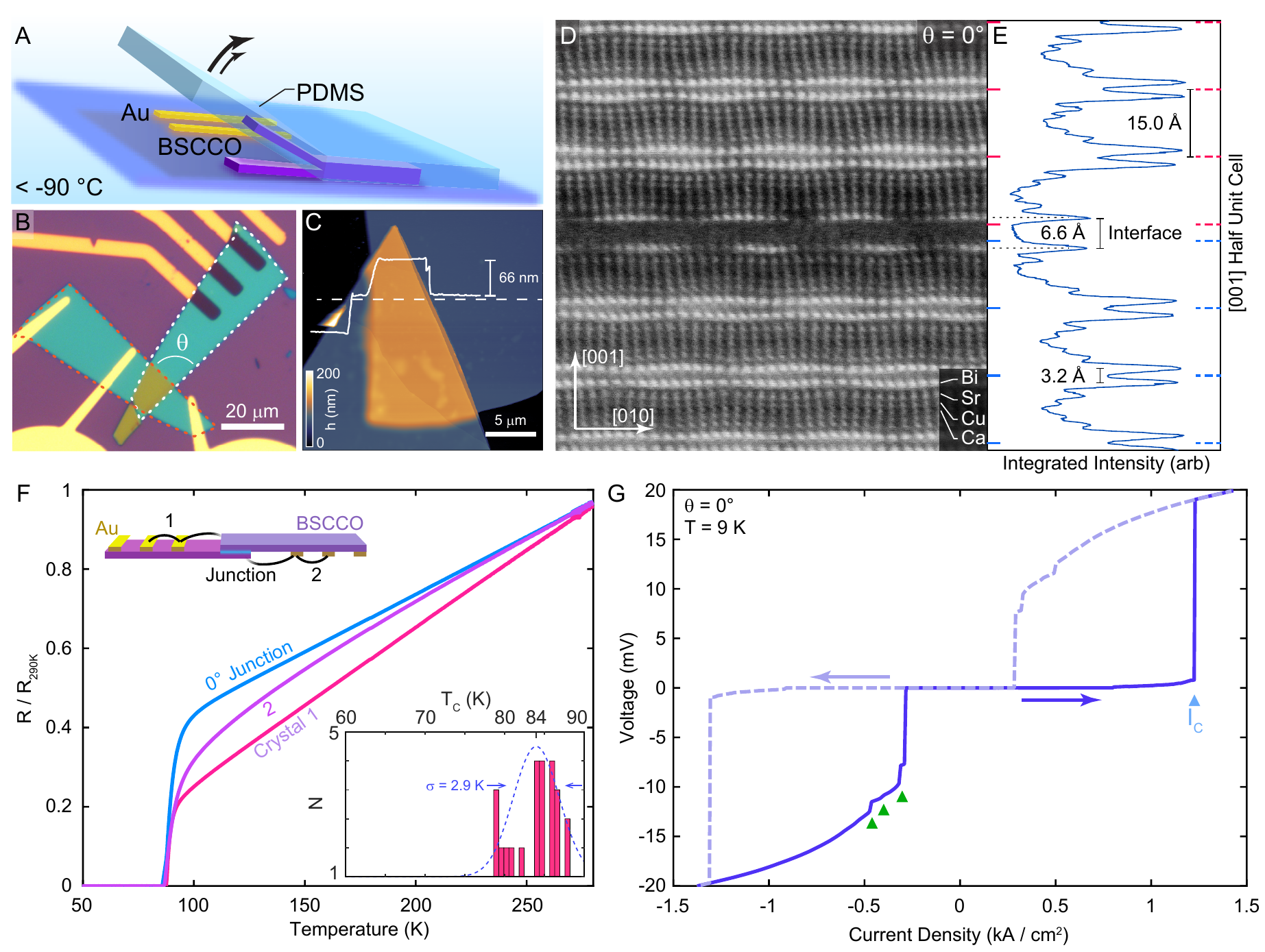}
		\setlength{\belowcaptionskip}{-20pt}
		\vspace{-0.4cm}		
		\caption
		{
			\textbf{Twist Josephson junctions with intrinsic junction quality. A.} Schematic of the key fabrication step, where a single BSCCO crystal is cleaved using PDMS below $-90^\circ$ C. \textbf{B.} Optical micrograph of a BSCCO twist junction. Dashes outline identical shapes of the two crystals. Corresponding schematic in \textbf{upper inset of F. C.} Atomic force microscope topography showing atomically flat interface. Line trace shows topography along dotted line. \textbf{D.} Cross-sectional scanning TEM image of junction at $\theta = 0^\circ$, showing crystalline order at the interface. Bright spots are columns of atoms which scatter electrons, the brightest of which are Bi. \textbf{E.} Integrated intensity at each layer. \textbf{F.} In-plane resistance in each bulk crystal \textit{vs} resistance through a twist junction between them, showing nearly identical junction $T_C$. \textbf{Lower Inset} shows $T_{C}$ distribution among all 24 JJs in the angle dependence analysis. \textbf{G.} $I-V$ curve for a $\theta = 0^\circ$ junction in both sweep directions (arrows). Blue triangle highlights $J_C$ comparable to intrinsic junctions. Green triangles highlight inelastic scattering features seen at the same voltages in intrinsic junctions \cite{IntrinsicStructure}.
		}
		\label{TwistDevice}
	\end{center}
\end{figure*}

We overcome these challenges by developing a cryogenic, solvent-free vdW transfer technique in pure argon using a liquid nitrogen-cooled stage kept $<-90^\circ$ C. We cleave an exfoliated BSCCO crystal into two copies between BiO planes, while thermally freezing out oxygen migration and other chemical processes at the surface (Figure 1A and SM Fig S1). One of the crystals is quickly rotated to the targeted twist angle, and re-assembled with the other. A Josephson junction forms in the overlapping region upon contact (Figure 1B and C). Two sets of electrical contacts, defined via stencil masks and evaporated on a -30$^\circ$ C cold stage \cite{BSCCOHall}, are pre-fabricated nearby before cleaving and placed on the top surface of the bottom crystal after re-assembly respectively. This contact geometry probes the twist junction while minimizing bulk crystal contributions (Fig. 1F upper inset). Additional details appear in SM Section S1.

Cryogenic handling in argon is critical to maintaining a pristine atomic interface without interfacial reconstruction and oxygen dopant changes. Figure 1D shows cross-sectional high-angle annular dark field (HAADF) scanning TEM image of a $\theta = 0^\circ$ junction. Crystalline order is well preserved at the interface along with structural supermodulations \cite{BSCCOStructure}.  We fabricated 24 devices with different twist angle $\theta$ between 0$^\circ$ to 180$^\circ$ with $T_C \geq 79$~K, and average $T_C$ of 84~K (Fig~1F lower inset), demonstrating high oxygen dopant uniformity even at the junction. Neither $T_C$ nor the normal-state conductivity are systematically correlated with $\theta$ (see Fig. S2). 

At $\theta = 0^\circ$, our devices exhibit electronic characteristics similar to single-crystal intrinsic junctions, demonstrating high interfacial quality of our JJs. Fig. 1G shows the $I$-$V$ curve measured with four terminals at temperature $T=9$~K. In this low temperature regime, the JJ exhibits a large hysteresis. As we increase current bias from a large negative value, the junction voltage $V$ first re-traps to the zero resistance state ($V = 0$) and then jumps to the resistive state at the critical current $I_C$, marked by the blue triangle. Upon reversing the bias current polarity (dashed line), the JJ's $I$-$V$ behavior is mirrored along $I=0$. Normalizing to junction area, we obtain a critical current density $J_C\approx1.2$~kA/cm$^2$ for this junction, similar to $J_C$ of intrinsic junctions \cite{IntrinsicIc}. We observe small voltage jumps on the retrapping side (green triangles) at the same voltages as inelastic tunneling features previously observed in intrinsic BSCCO JJs \cite{IntrinsicStructure}. These observations indicate that our $\theta=0^\circ$ JJ reaches electronic quality comparable to intrinsic JJ in single-crystal BSCCO. 

\begin{figure*}[!tb]
	\vspace{-0.5cm}
	\begin{center}
		\includegraphics[width=1\linewidth]{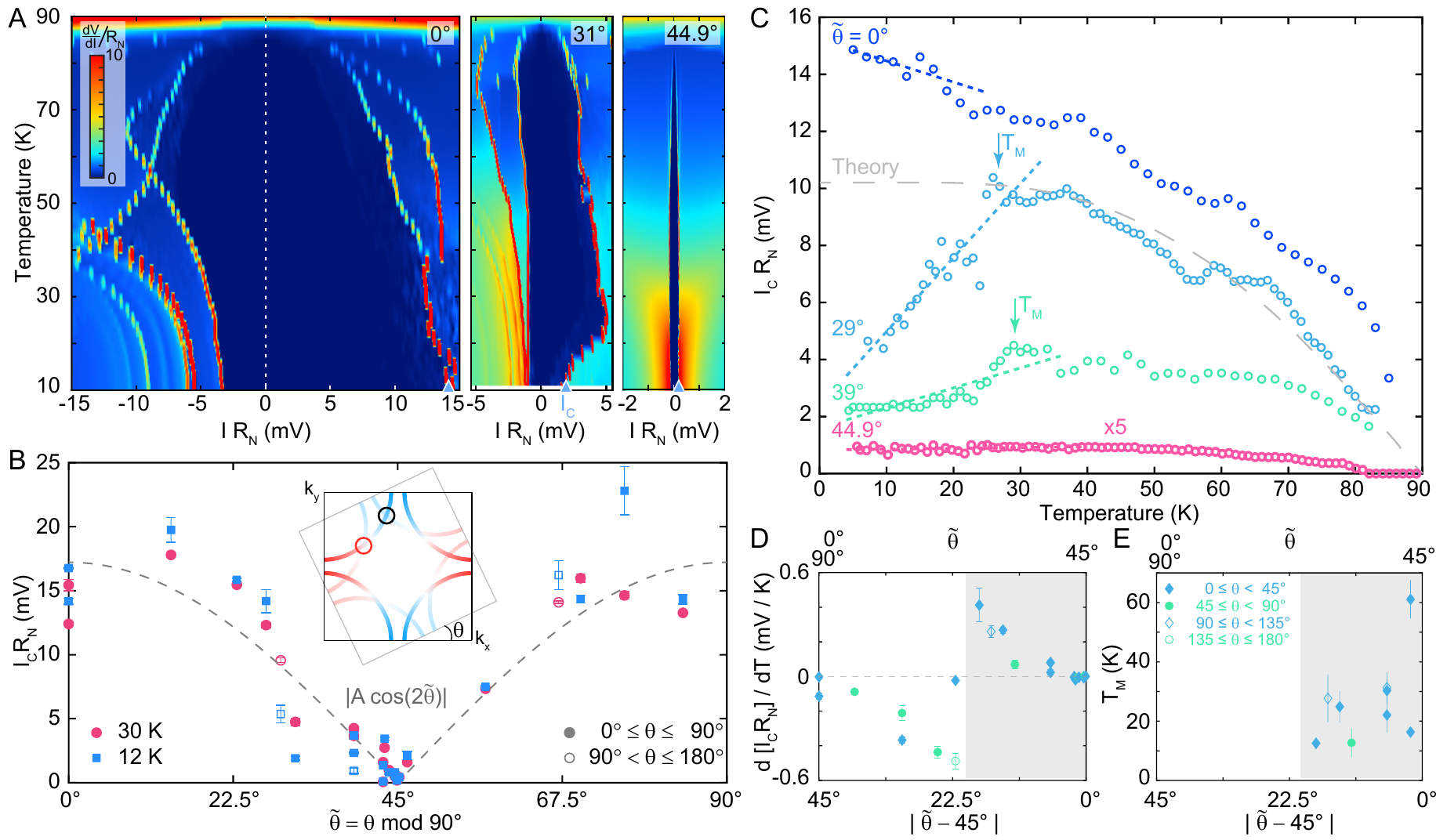}
		\setlength{\belowcaptionskip}{-20pt}
		\vspace{-0.4cm}		
		\caption
		{
			\textbf{$\mathbf{d}$-wave SOP symmetry revealed by supercurrent tunneling. A.} Normalized differential resistance $[dV/dI] / R_N$ \textit{vs.} characteristic voltage $IR_N$ and temperature $T$. Current is swept to the right. Blue arrows highlight $I_CR_N$. \textbf{B. } Angular dependence of $I_CR_N$ for all devices at $30$ and $12$~K. The points follow the $|\cos(2\tilde{\theta})|$ curve predicted for nearly incoherent tunneling between $d$-wave superconductors \cite{TwistTheoryRev}. \textbf{Inset:} Schematic diagram of the Fermi surface of both crystals, with sign and magnitude of superconducting gap $\Delta(\vec{k})$ superposed in color. At $\theta > 10^\circ$, Fermi surfaces intersect at two points per quadrant (circles) with different relative signs of SOPs. \textbf{C.} Temperature dependence of the critical current for select devices. Dotted lines are linear fits to the low temperature data. $T_M$ is temperature where $I_C$ is maximal. Grey theory line shows expected $I_CR_N(T)$ behavior (see Section S5). \textbf{D.} The slope of the low temperature linear fit, $d(I_CR_N)/dT$. \textbf{E.} $T_M$ as a function of angle $\tilde{\theta}$.
		}
		\label{TwistAngleDep}
	\end{center}
\end{figure*}

To compare transport characteristics of different twisted JJs, we normalized the bias current $I$ with the junction normal resistance $R_N$. Since $I_C$ and $R_N^{-1}$ are proportional to the area of the junction, the product $I_CR_N$ is independent of junction area. Figure 2A shows the normalized dynamic resistance $[dV/dI]/R_N$ as a function of $T$ and $I R_N$ at $\theta=0^\circ$, $31^\circ$, and $44.9^\circ$, respectively (similar data for all 24 JJs studied are shown in Fig. S3). Several features are apparent in these data sets. First, as the current sweeps from left to right, on the retrapping side ($IR_N<0$), constant-voltage inelastic tunneling features appear (green triangles in Fig. 1G) in arcs of constant $V$. Next, on the switching side ($IR_N>0$), both $V$ and $dV/dI$ jump at critical current $I_C$, which depends on $T$. The detailed behavior of $I_C(T)R_N$ depends on $\theta$, as we will detail below. Finally, we find the hysteresis of JJs to be reduced in the high temperature regime as $I_C(T)R_N$ decreases.

Analyzing $I_CR_N$ for all 24 devices with $\theta$ between 0$^\circ$ and 180$^\circ$, we find that the magnitude of $I_CR_N$ becomes smaller closer to 45$^\circ$ and 135$^\circ$ where the JJs also appear less hysteretic. In Figure 2B, we plot $I_CR_N$ at two representative temperatures 12~K and 30~K as a function of a new variable $\tilde{\theta}=\theta$ (mod $\pi/2$). We observe that $I_CR_N(\tilde{\theta})$ follows $|\cos(2\tilde{\theta})|$, which is expected for somewhat incoherent Cooper pair tunneling between $d$-wave superconductors~\cite{TwistTheoryRev}. Similar behavior is seen in the angular dependence of junction voltage just above the critical current $V(I_C)$ (see Fig. S5). In conventional tunnelling JJs,  $eV(I_C)\approx 2\Delta$ \cite{Barone}. 
	
The temperature dependence of Josephson coupling in our twisted junctions provides further insight into the pairing symmetry of the Cooper pairs in BSCCO. Figure 2C shows $I_C(T)R_N$ for several representative JJs with different $\tilde{\theta}$. For $\tilde\theta \sim 0$, we find that $I_C(T)R_N$ monotonically decreases as $T$ increases, approximately following the  theory curve for nearly incoherent tunneling between $d$-wave superconductors (dashed line, see SM Section S5). As $\tilde{\theta}$ increases, however, a surprising non-monotonic behavior of $I_C(T)R_N$ appears. For example, for $\tilde{\theta}=29^\circ$ and $39^\circ$ in Fig~2C, $I_C(T)R_N$ increases alongside $T$, reaching a maximum value at $T=T_M$ and then decreases as $T$ approaches $T_C$. More quantitative analysis can be found in Figure 2D and Figure 2E, where we plot the low temperature slope $d(I_CR_N)/dT$ (dotted lines in Fig~2C and S4) and observed $T_M$. These plots show a non-monotonic behavior of $I_CR_N(T)$, signaled by the positive slope of $I_CR_N(T)$ at low temperatures with finite $T_M$, appearing within $|\tilde{\theta}-\pi/4|<\pi/8$.

The strong $\tilde \theta$ dependence of the non-monotonic $I_C(T)R_N$ in Fig 2D and E points to SOP $d$-wave symmetry as its origin. For this explanation, we consider a gap function $\Delta_{1,2} (\mathbf k)$ superimposed on top of the Fermi surface $E_F^{1,2}(\mathbf{k})$ \cite{hashimoto2014}, where $\mathbf{k}$ is the in-plane Cooper pair wavevector in the first Brillouin zone and the index 1 or 2 denotes the top and bottom layer of BSCCO, respectively. In the twisted JJ, $E_F^{1}(\mathbf{k})$ and $E_F^{2}(\mathbf{k})$ are rotated relative to each other by angle $\theta$ (Fig 2B inset). At $\theta \approx 0$, $E_F^1(\mathbf{k})$ and $E_F^2(\mathbf{k})$ overlaps almost completely and $\Delta_1(\mathbf{k}) \Delta_2(\mathbf{k}) > 0$, yielding a uniformly positive contribution to critical current for coherent tunneling \cite{TwistTheoryRev}.
As $\theta$ increases to $\sim 10^\circ$, the Fermi surfaces overlap at two points per quadrant in $\mathbf k$-space, but with opposite phase difference between layers, yielding nodal and anti-nodal contribution where $\Delta_1(\mathbf{k})\Delta_2(\mathbf{k})$ alternate in sign. Since the supercurrent from these two components carry the opposite sign, their contributions to the total critical current compete. 
As the gap in the nodal region is much smaller than the antinodal one, non-monotonic temperature dependent $I_C(T)$ is expected for $\tilde{\theta} \approx \pi/8$ where the competition is strongest (see SM Section S5 and Ref.\ \cite{Plugge2021} for more quantitative discussion). 
We also note that near $\tilde{\theta} \approx \pi/4$, the JJ coupling is strongly suppressed but remains non-zero. For the $\theta = (44.9\pm.1)^\circ$ junction, Josephson critical current can be measured up to 79 K with $I_CR_N$ about two orders of magnitude smaller than the $0^\circ$ value.

\begin{figure*}[!tb]
	\vspace{-0.0cm}
	\begin{center}
		\includegraphics[width=1\linewidth]{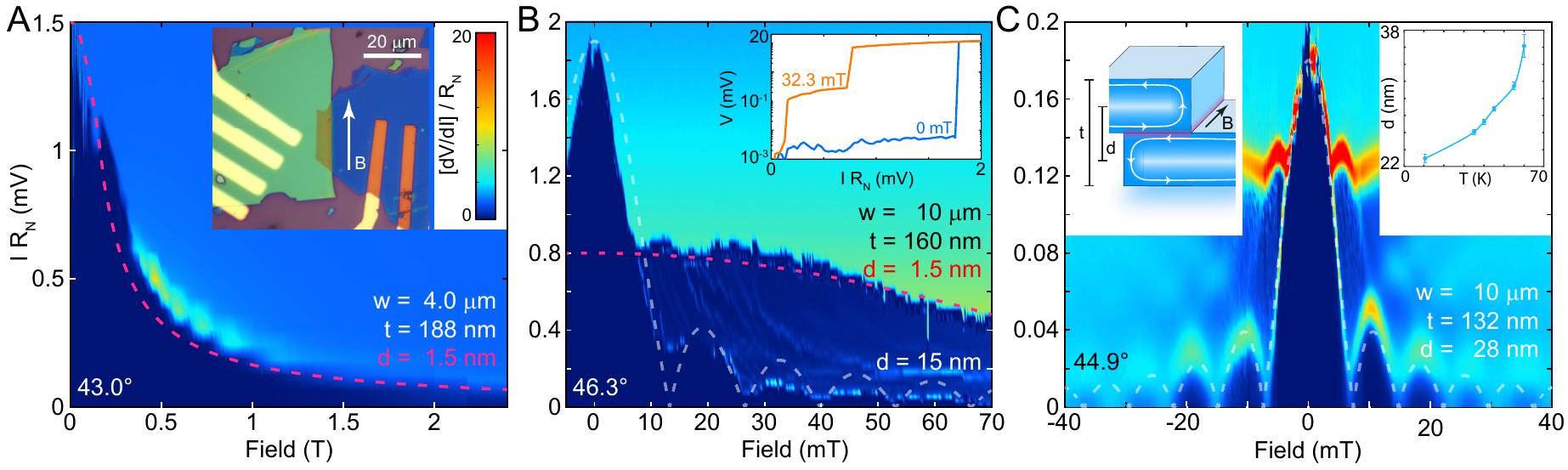}
		\setlength{\belowcaptionskip}{-20pt}
		\vspace{-0.4cm}		
		\caption
		{
			\textbf{Fraunhofer pattern near $\mathbf{45^\circ}$.} $dV/dI$ response to in-plane magnetic field $B_{\parallel}$ at three angles with width $w$ perpendicular to $\vec{B_\parallel}$, measured from microscope images, and thickness $t$ measured via atomic force microscopy. \textbf{A.} $43.0^\circ$ JJ. Red dashed lines show the Fraunhofer pattern envelope with effective magnetic thickness $d=1.5$~nm expected in intrinsic BSCCO JJ. \textbf{Inset} shows device optical photo. \textbf{B.} $46.3^\circ$ JJ. As $\theta$ approaches $45^{\circ}$, an additional Fraunhofer pattern with shorter magnetic field periodicity appears (grey dashed line), with an estimated $d=15$~nm, coexisting with the $d = 1.5$~nm pattern (red). \textbf{Inset} shows $I$-$V$ characteristics at two different $B_\parallel$ values, showing two jumps in $V$. \textbf{C. } Response in a JJ at 44.9$^\circ$. One well-developed Fraunhofer pattern corresponding to $d=28$~nm ($14$ nm for a second-harmonic CPR) appears, as indicated by the grey dashed line. \textbf{Left inset} shows junction schematic. The Meissner currents (white lines) in the flakes affect the phase difference at the twist junction, enhancing its effective thickness $d$ for the magnetic flux \cite{Barone} (see SM Section S9). 
			The \textbf{right inset} shows experimentally evaluated $d$ at different temperatures on the $44.9^\circ$ JJ. All values of $d$ assumes conventional (1st order) Josephson coupling.
		}
		\label{TwistFraunhofer}
	\end{center}
\end{figure*}

The origin of the finite supercurrent near $45^\circ$ is encoded in the Josephson current-phase relation (CPR) ~\cite{canHightemperatureTopologicalSuperconductivity2021,Goldobin,sigrist1998}. At $\tilde\theta =45^\circ$, the JJ coupling lacks the conventional direct Cooper pair tunneling term $I_c\sim \sin\varphi$, where $\varphi$ is the SOP's phase difference across the interface, due to the maximally mismatched SOP across the twisted interface. The supercurrent must then tunnel through a second-order mechanism corresponding to co-tunneling of Cooper pairs, which is predicted to support an interfacial SOP with emergent $d_{x^2-y^2} + id_{xy}$ symmetry \cite{canHightemperatureTopologicalSuperconductivity2021, GeneralTwist, sigrist1998}. The Josephson CPR of this unusual SOP develops a strong second-order harmonic $I_c^{(2)}\sim \sin2\varphi$ \cite{sigrist1998, canHightemperatureTopologicalSuperconductivity2021, Goldobin}, whose signature can be experimentally probed by measuring the in-plane magnetic interference ('Fraunhofer') pattern or microwave induced Shapiro steps in the $I$-$V$ characteristic, which are both sensitive to the $4e$ charge of co-tunneling Cooper pairs across the junction~\cite{Goldobin, sigrist1998, GeneralTwist}.

Figure 3 shows Fraunhofer interference patterns (FIP) obtained at three different angles by applying parallel magnetic field $B_\parallel$. For the $43.0^\circ$ junction (Fig. 3A), we observe $I_C$ suppression without clear oscillatory behavior, resembling intrinsic JJ FIP~\cite{IntrinsicIc}, where the junction effective magnetic thickness $d$ is equal to the vdW layer spacing $s = 1.5$ nm \cite{TheoryFraunhofer}. Very close to $\tilde\theta = \pi/4$, however, well-defined FIP oscillations appear. Fig. 3C shows $\theta=44.9^\circ$ JJ exhibit $I_C(B_\parallel)$ oscillation with period about 20 times shorter than that expected for intrinsic junctions. At an intermediate angle $\theta=46.3^\circ$, (Fig. 3B), the long and short-period oscillations appear to coexist. The reduction in FIP period implies an increase of $d$ due to field-induced currents extending into the crystal bulk~\cite{Barone} (see SM Section S9). The ratio $d/t$, where $t$ is the actual junction thickness, depends only on the properties of the crystal and junction geometry, and should not depend strongly on the twist angle (see SM Section S9). Intriguingly, for the devices in figure 3B and C which share similar geometry, we obtain $d/t \approx 0.1$ for both devices only if we assume that the $44.9^\circ$ junction is coupled purely through the co-tunneling process with a doubled FIP period. 

\begin{figure*}[!ht]
	\vspace{-0.5cm}
	\begin{center}
		\includegraphics[width=0.66\linewidth]{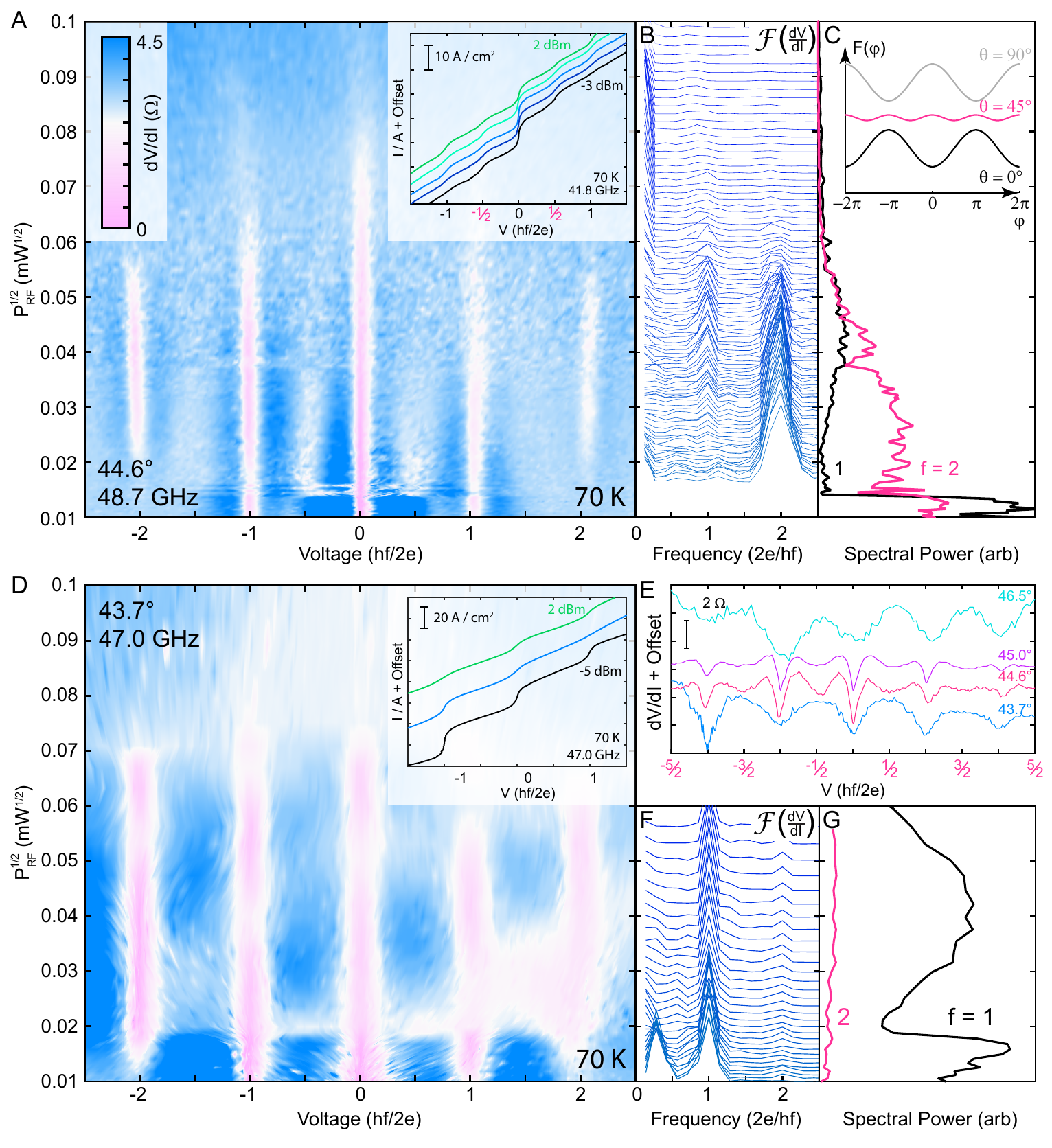}
		\setlength{\belowcaptionskip}{-20pt}
		\vspace{-0.4cm}		
		\caption
		{
			\textbf{Half-integer Shapiro steps emerge close to $\mathbf{\theta = 45^\circ}$}. \textbf{A.} $dV/dI$ as a function of voltage across the junction and microwave illumination power $P_{RF}^{1/2}$ at $70$ K. $dV/dI$ dips (white, pink) correspond to Shapiro steps. \textbf{Inset} shows $I$-$V$ characteristic with half-integer Shapiro steps. \textbf{B.} and \textbf{C.} show the Fourier transform of $dV/dI (V)$ and the spectral power at $\nu_f = 1$ and $2 \cdot 2e/hf$. \textbf{Inset} shows schematic of the junction free energy $F$ vs Josephson phase $\varphi$ as twist angle changes. At $45^\circ$, the second harmonic dominates the current-phase relation. \textbf{D, F and G} shows similar data for a $43.7^\circ$ device, where only integer Shapiro steps appear. \textbf{E.} shows representative $dV/dI$ for all four devices with Shapiro step measurements at different twist angles. Half integer Shapiro steps are only observed in junctions closest to $45^\circ$
		}
		\label{TwistShapiro}
	\end{center}
\end{figure*}

The presence of higher order harmonics in the CPR near $45^\circ$ are further revealed by measuring Shapiro steps in $I$-$V$ under microwave illumination of frequency $f$ (inset of Fig. 4A and D) \cite{Goldobin, GeneralTwist}. In conventional JJs where the CPR is dominated by the first harmonic of $\varphi$, Shapiro steps appear as plateaus of constant voltage whenever $V$ approaches $n \cdot hf/2e$, where $n$ is an integer. We observe these conventional integer Shapiro steps in the JJs substantially away from $\tilde\theta =\pi/4$, as shown in the $\theta =$ 43.7$^\circ$ device (Fig. 4D). Consistent with the FIP discussed above, the experimentally observed Shapiro steps also show signatures of the second harmonic CPR as $\tilde\theta $ approaches $\pi/4$ (Fig. 4E); specifically, when our devices are within $(45\pm1)^\circ$, additional steps at \textit{half-integer} $n$ appear. As shown in Fig. 4A ($\theta=44.6^\circ$ as an example), a series of $dV/dI$ dips which correspond to the steps in $I$-$V$ appear at both integer and half-integer $n$ across a wide range of microwave power. 

The Fourier components of $dV/dI (V)$ shows the relative strengths of integer and half-integer Shapiro steps to estimate our junction CPR. For the conventional Shapiro steps appearing in the 43.7$^\circ$ device, the Fourier transform shows dominant spectral power for the first harmonic $\nu_1 = 2e/hf$ (Fig. 4F and G). However, for the devices exhibiting half-integer Shapiro steps (e.g.\ the $44.6^\circ$ device at low microwave power), the Fourier transform is dominated by the second harmonic $\nu_2 = 4e/hf$. The corresponding $dV/dI$ shows dips of similar strength at half integer and integer steps, indicating that the co-tunneling of Cooper pairs dominates over the conventional Josephson coupling close to $45^\circ$. The presence of the dominant second harmonic CPR, demonstrated in $(45\pm1)^\circ$ JJs, establishes the unconventional nature of the interfacial high-temperature superconductivity expected to support a topological superconducting phase which spontaneously breaks time reversal symmetry \cite{canHightemperatureTopologicalSuperconductivity2021, GeneralTwist, sigrist1998}.

\bibliography{references.bib}

\bibliographystyle{ieeetr}

\section*{Acknowledgments}
The experiments were supported by the NSF (DMR-1809188 and DMR-1922172). PK acknowledge the support from the U.S. Department of Defense (DOD) Vannevar Bush Faculty Fellowship N00014-18-1-2877. Work at UBC was supported by NSERC and CFREF. Stencil masks were fabricated at Harvard CNS, a part of National Nanotechnology Coordinated Infrastructure, NSF 1541959. NP acknowledges the Deutsche Forschungsgemeinschaft (DFG-452128813) for partial support with the project. P.A.V.\ is supported by a Rutgers Center for Materials Theory Postdoctoral Fellowship and
J.H.P.\ is partially supported by the Air Force Office of Scientific Research under Grant No.~FA9550-20-1-0136, NSF CAREER Grant No. DMR-1941569, and  the Alfred P. Sloan Foundation through a Sloan Research Fellowship.
P.A.V.\ and J.H.P.\ acknowledge the Aspen Center for Physics where part of this work was performed, which is supported by National Science Foundation grant PHY-1607611. This work was partially supported by a grant from the Simons Foundation (P.A.V.). The work at BNL was supported by the US Department of Energy, oﬃce of Basic Energy Sciences, contract no. DOE-sc0012704. The authors are grateful for sample shipping coordination with Joon Young Park, and fruitful discussion with D. Kwabena Bediako, Ken S. Burch, Srivatsan Chakram, Gil-Ho Lee, R\'egis M\'elin, and Justin Wilson.

\section*{Competing Interests}
Authors declare that they have no competing interests.

\section*{Author Contributions}
SYF.Z., N.P. and P.K. conceived and designed the experiment; SYF.Z. and N.P. developed the air-sensitive cryogenic stacking technique; SYF.Z., X.C. and N.P. performed the experiments. H.Y. and R.E. performed the STEM experiment. R.Z. and G.D.G. provided the crystals. P.A.V. and J.P. performed theoretical analysis of the Fraunhofer patterns and contributed to the analysis of the critical current. M.F., S.P. and T.T. performed theoretical analysis of the critical current.  SYF.Z., N.P., X.C. and P.K. analyzed the data and wrote the manuscript with contribution from P.A.V., J.H.P, M.F., and Y.R.

\section*{Supplementary Materials}
Methods, Detailed Analysis, and Additional Data\\
Figs. S1 to S9\\
Tables S1\\
Additional References 33 - 45.


\newpage

\renewcommand{\thefigure}{S\arabic{figure}} 
\renewcommand{\thesection}{S\arabic{section}} 
\renewcommand{\thetable}{S\arabic{table}} 
\renewcommand{\theequation}{S\arabic{equation}}
\setcounter{figure}{0} 

\section*{Supplementary Materials: Emergent Interfacial Superconductivity between Twisted Cuprate Superconductors}

\section{Sample Fabrication Method}
We have developed a novel, all-dry, cryogenic pick-and-place technique to fabricate our Josephson junctions. Our technique allows us to cleave a pair of fresh surfaces of BSCCO from one pre-exfoliated single crystal, and then quickly stack the crystals together forming the twist junction. Oxygen dopants are conveniently frozen alongside any chemical degradation processes below -90$^\circ$~C \cite{fratiniScalefreeStructuralOrganization2010}, preserving interfacial crystallinity and superconductivity. The entire procedure can be cleanly performed in an argon glovebox without solvents, as our transfer polymer does not melt at the final drop-off step. The twist angle can also be accurately controlled and measured to $0.1^\circ$ resolution using optical microscopy, since the junction is made starting from one single crystal. This cryogenic pick-and-place technique should be generally applicable to any air- and heat-sensitive material. 

All dry vdW pickup techniques \cite{LeiPickup, PCPickup} relies on temperature to control adhesion to a polymeric transfer stamp \cite{ShapeMemoryPolymers, TransferReview}. We use poly(dimethylsiloxane) (PDMS) to decrease the glass transition temperature $T_g$ to about $-100^\circ$ C \cite{PDMSTg}, which is accessible to a liquid nitrogen cooled stage in a pure argon environment. We are careful to set the substrate temperature above the frost-point of our glovebox, where ice visibly deposits on our substrates. This is carefully kept below $-100^\circ$ C, corresponding to roughly 10 parts per billion (ppb) of H$_2$O by volume \cite{GoffGratch}. Finally, PDMS freely releases vdW crystals onto the substrate at $-35^\circ$ C without melting \cite{PDMSPickup}. 

We prepare our PDMS using Dow Corning Sylgard 184, mixed to 10:1 base:curing agent ratio. The mixture is poured onto a clean glass slide to form a flat layer 1mm thick, vacuum degassed, and oven baked overnight at 65$^\circ$ C. The cured PDMS is then cut into about $2 \times 2$ mm squares, placed onto a glass slide, and baked to 300 $^\circ$C for 15 minutes for adhesion. 

\begin{figure*}[ht!]
		\begin{center}
			\includegraphics[width=1\linewidth]{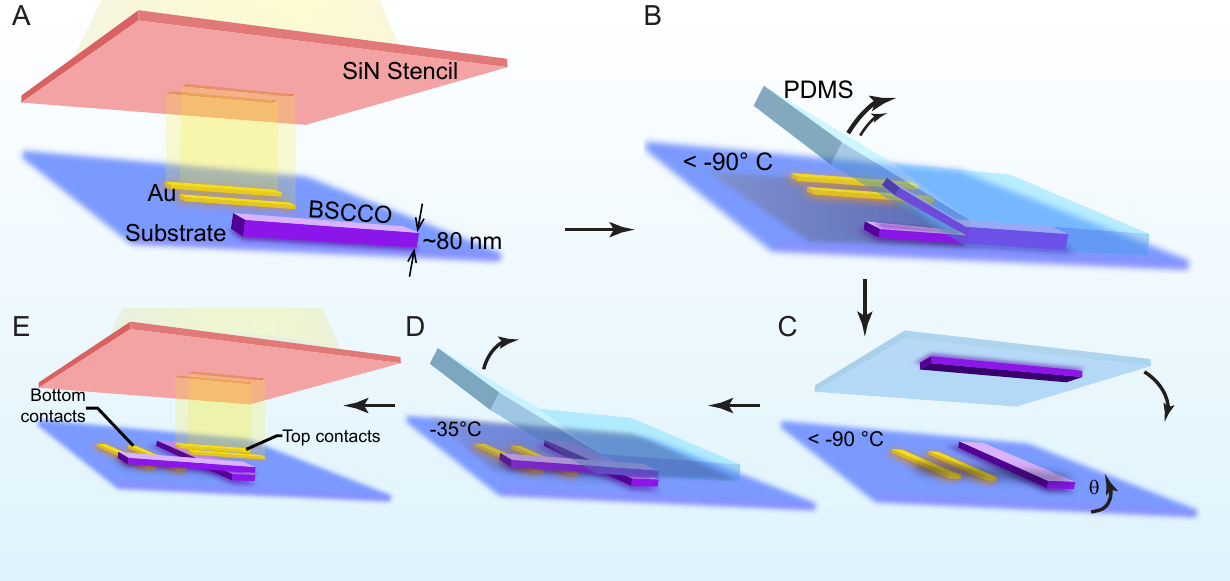}
			\setlength{\belowcaptionskip}{-20pt}
			\vspace{-0.4cm}		
			\caption
			{
				\textbf{Fabrication Process. A.} BSCCO is identified and gold contacts pre-evaporated next to it. \textbf{B.} Cold PDMS is quickly pulled away, cleaving the crystal. \textbf{C.} Substrate is quickly rotated by $\theta$, and top crystal is quickly re-assembled with the bottom. \textbf{D.} Assembly is warmed to -35$^\circ$ C, and PDMS slowly removed. \textbf{E.} Top contacts on bottom crystal deposited. 
			}
			\label{Fig1}
		\end{center}
	\end{figure*}

We prepare our silicon substrates by first baking them to 300$^\circ$ C overnight, and then cleaning them with oxygen plasma for 1 minute. We then exfoliate optimally doped Bi$_2$Sr$_2$CaCu$_2$O$_{8+x}$ on cooled substrates at -35$^\circ$ C using 3M Scotch tape.

We then identify a flat, near optimally-doped BSCCO crystal roughly 80~nm thick, and evaporate gold contacts next to it using a stencil mask technique \cite{BSCCOHall}, on a Peltier-cooled evaporator sample stage (-30$^\circ$~C) (Figure S1A). We then cool the substrate to -80$^\circ$~C, touch a small corner of a PDMS stamp to the BSCCO, and cool the assembly below -90$^\circ$ C. Once cold, we quickly \cite{PDMSSpeed} pull the PDMS off. The competition of adhesion forces between BSCCO, the substrate and cold PDMS often cleaves the crystal along an atomically flat plane between BiO planes (Figure S1B). We rapidly rotate the substrate by an angle $\theta$, and place the upper BSCCO crystal on top of both the lower crystal and gold contacts (Figure S1C). We find that the time between cleaving and reassembly strongly influences junction quality, and should be kept below 3 minutes. Next, the substrate is heated to -35$^\circ$ C and the PDMS slowly \cite{PDMSSpeed} removed (Figure S1D). Finally, a second set of gold contacts is evaporated onto the \textit{top} surface of the \textit{bottom} BSCCO crystal using a stencil mask (Figure S1E), which minimizes resistance contribution from intrinsic Josephson junctions along the c-axis in each bulk crystal. We emphasize that the BSCCO crystals were kept at or below room temperature, and away from air throughout the fabrication process. Time between fabrication and measurements are kept as short as possible.

\section{Measurement Method}
All electrical measurements were performed in 4 terminal geometry to eliminate contact resistances. $dV/dI$ measurements are performed by adding AC ($15 < f < 150$ Hz) and DC voltages generated by a Stanford Research Systems SR830 lock-in amplifier (with $1/1000 \times$ voltage divider) and Keithley 2400 SourceMeter respectively, using a toroidal transformer. The voltage output passes through a large resistor to form a current source. The voltage across the junction is amplified with a SR570 low-noise preamp, and measured with a Agilent 34401A multimeter and SR830 lock-in amplifier. Cryogenic temperatures are reached using liquid helium flow cryostats. 

Shapiro step measurements are performed by generating a $f_{RF} < 50$ GHz microwave signal (Keysight E8257D) and guiding it to the sample through a low-loss semi-rigid coaxial cable with a 1.85 mm air dielectric connector (Pasternack PE3C6490). It is epoxy-set into a KF-25 adapter at the cryostat wall (Torr-Seal). The coax shield is cut about 3 mm shorter than the center wire, forming an improvised antenna a few mm above the sample substrate. Care was taken to minimize coax cable bending.

The AFM topography data is taken after the electrical measurement, in a Park AFM system in non-contact mode.  

Cross-sectional STEM specimen was prepared by Ga ion milling in a focused ion beam (FIB) (Helios G4, Thermo Fisher Scientific). The surface region of the specimen that was damaged during the FIB process was removed by low-energy Ar ion milling system (NanoMill 1040, Fischione) to improve image quality. A spherical aberration corrected STEM (JEM-ARM 200F, Jeol) with the acceleration voltage of 200 kV was used to obtain atomic resolution image. Inner collection angle of 68 mrad was used for HAADF STEM imaging.


\section{List of Devices for Angle-Dependent $I_CR_N$ Analysis}
Here we list all devices used in the angle-dependent $I_CR_N$ analysis. We have excluded non-superconducting devices and those with less than 2 contacts on each side of the Josephson junction, which is necessary for a 4-point measurement. To keep doping levels consistent, we have also eliminated junctions with superconducting transition temperature below 79 K. We take $T_C$ to be the temperature where the junction resistance falls to within 1\% of the resistance value at $90$~K, at the zero current bias limit. 

$R_N$ is extracted just under $T_C$ at moderate bias far above the junction critical current, but below the in-plane transport critical current, in order to minimize in-plane contributions. 

\begin{table}[h!]
	\begin{center}
		\caption{\textbf{Summary of device transport characteristics.}}
		\label{DevCharacteristics}
		\begin{tabular}{r|r|r|r|r|r} 
			$\theta$ 	& Area 		    & $T_C$ 	& $R_N$ 	    & $R_N^{-1} /$ Area 	& $I_C(30K) / $ Area     \\
			degrees 	& $\mu$ m$^2$ 	& K 	    & $\Omega$      &mS/$\mu$ m$^2$       	& A $/$ cm$^2$ \\
			\hline
            0	    & 376	& 86.7	& 3.6	& 0.743	& 1220  \\
            0	    & 1393	& 85	& 1.9	& 0.388	& 577  \\
            14	    & 255	& 79.1	& 9.3	& 0.422	& 927  \\
            23  	& 522	& 88.1	& 3.0	& 0.639	& 1050 \\
            27  	& 497	& 85.9	& 6.9	& 0.294	& 326  \\
            119	    & 736	& 81	& 11.4	& 0.119	& 55 \\
            31  	& 417	& 86	& 3.5	& 0.685	& 106  \\
            39  	& 446	& 85	& 2.0	& 1.132	& 250  \\
            39  	& 737	& 86	& 3.0	& 0.454	& 200  \\
            129	    & 144	& 84	& 8.8	& 0.787	& 140  \\
            43  	& 111	& 86.3	& 13.7	& 0.658	& 90  \\
            43  	& 170	& 79	& 7.7	& 0.764	& 14  \\
            43.2	& 77	& 85	& 16.6	& 0.782	& 222  \\
            43.8	& 568	& 88	& 1.8	& 0.978	& 119  \\
            44.6	& 151	& 80	& 3.2	& 2.083	& 116  \\
            44.6	& 294	& 84.4	& 6.2	& 0.546	& 45  \\
            44.9	& 123	& 79	& 4.0	& 2.022	& 38 \\
            45.2	& 182	& 84	& 5.9	& 0.931	& 32  \\
            46.3	& 130	& 84	& 2.9	& 2.643	& 412  \\
            57	    & 374	& 79.6	& 10.8	& 0.248	& 170  \\
            157	    & 148	& 86	& 10.6	& 0.638	& 1250  \\
            70  	& 293	& 86.7	& 8.6	& 0.397	& 671  \\
            76	    & 227	& 82	& 17.5	& 0.252	& 442  \\
            84  	& 296	& 84.1	& 5.9	& 0.573	& 848  \\
		\end{tabular}
	\end{center}
\end{table}

\begin{figure}[h!]
	\vspace{-0.5cm}
	\begin{center}
		\includegraphics[width=0.5\linewidth]{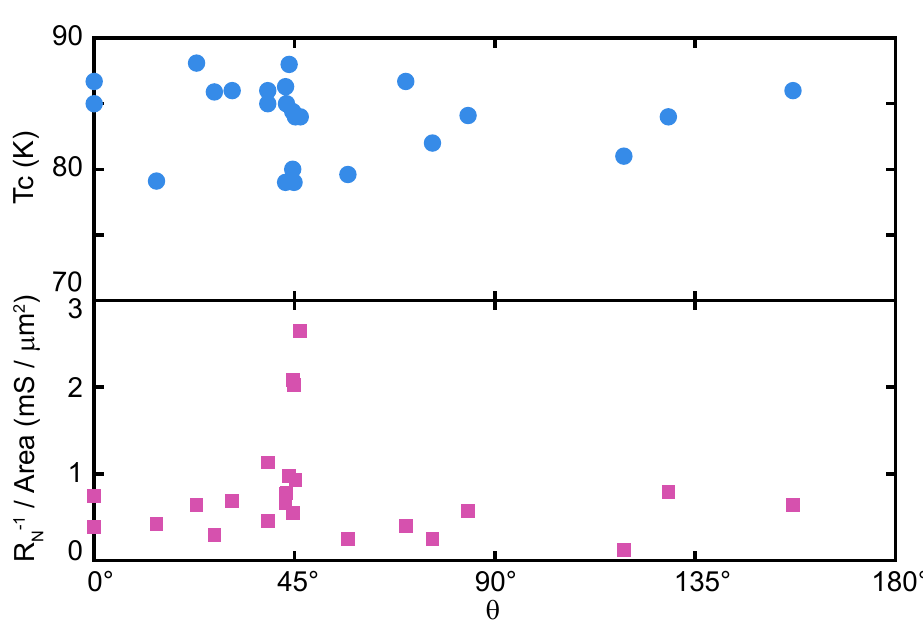}
		\setlength{\belowcaptionskip}{-20pt}
		\vspace{-0.4cm}		
		\caption
		{
			\textbf{Angular dependence of sample quality}. Junction $T_C$ and $R_N^{-1}$ / Area vs $\theta$ respectively, showing no systematic angular dependence of junction transport quality. 
		}
		\label{S2}
	\end{center}
\end{figure}

Figure S2 and Table S1 shows that while each device is somewhat unique, there is no systematic angular dependence on the junction $T_C$ or conductivity $R_N^{-1} / A$. The angular variation of $I_CR_N$ shown in figure 2 (main text) is due to intrinsic effects near $45^\circ$ rather than extrinsic differences in junction transparency or quality. Some devices near 45$^\circ$ are made with a mechanical jig to increase $\theta$ accuracy, which also reduced stacking time, and increasing the device quality.

\section{dV/dI Data for All Samples}
Figure S3 shows the dV/dI color plots for all devices, in the same format as shown in Fig 2A in the main text. Note that in most devices, only a single voltage jump is visible, implying no additional intrinsic Josephson junctions intruded the current path between voltage leads. Figure S4 shows the $I_CR_N(T)$ for all devices, in the same format as shown in Fig 2C in the main text.
\begin{figure*}[h!]
	
	\begin{center}
	\vspace{-3cm}
		\includegraphics[width=0.87\linewidth]{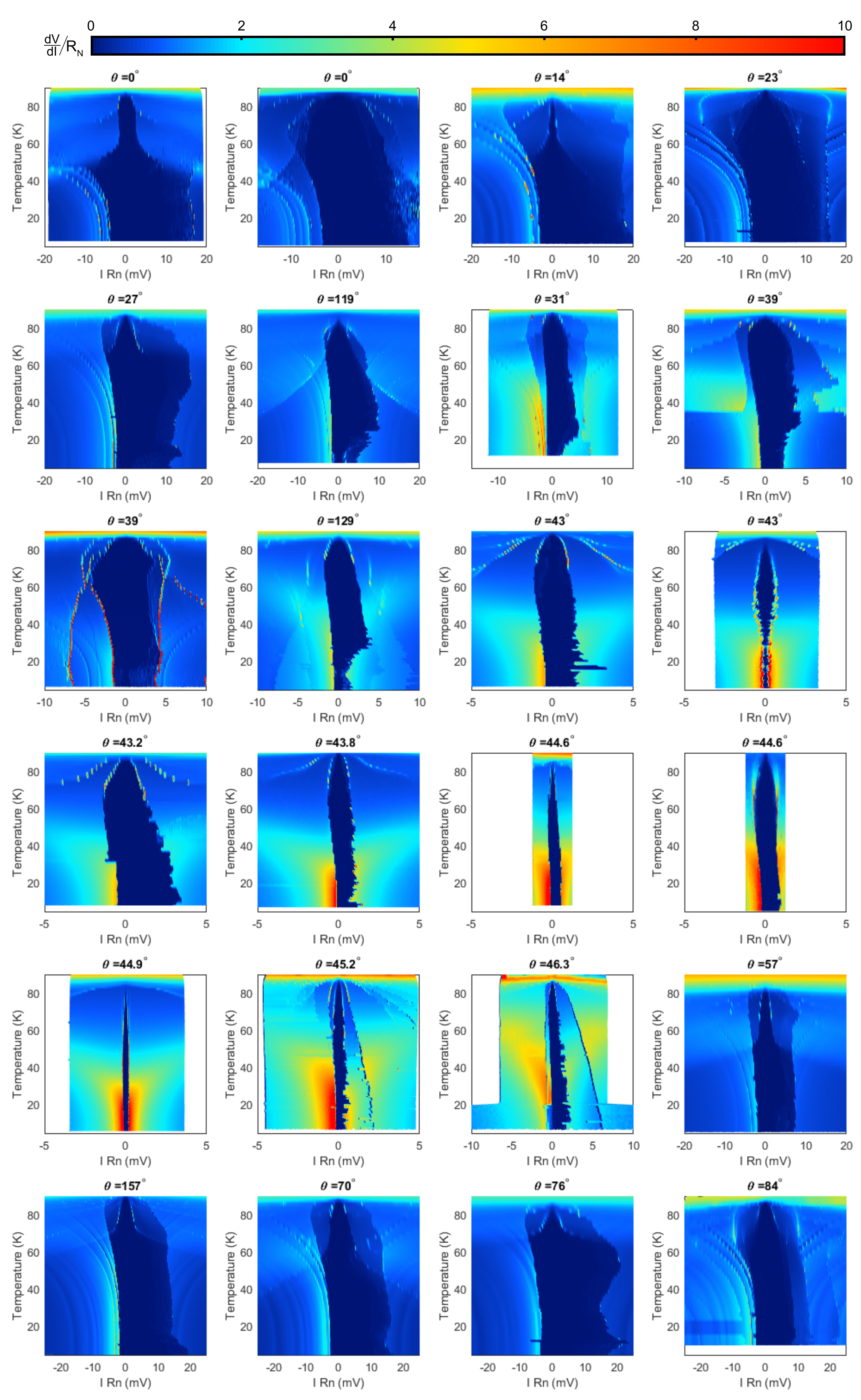}
		\setlength{\belowcaptionskip}{-20pt}
		\vspace{-0.4cm}		
		\caption
		{
			\textbf{[$dV/dI]/R_N$ for all junctions in the angle-dependence}. 
		}
		\label{S3}
	\end{center}
\end{figure*}

\begin{figure*}[h!]
	\vspace{-0.5cm}
	\begin{center}
		\includegraphics[width=0.8\linewidth]{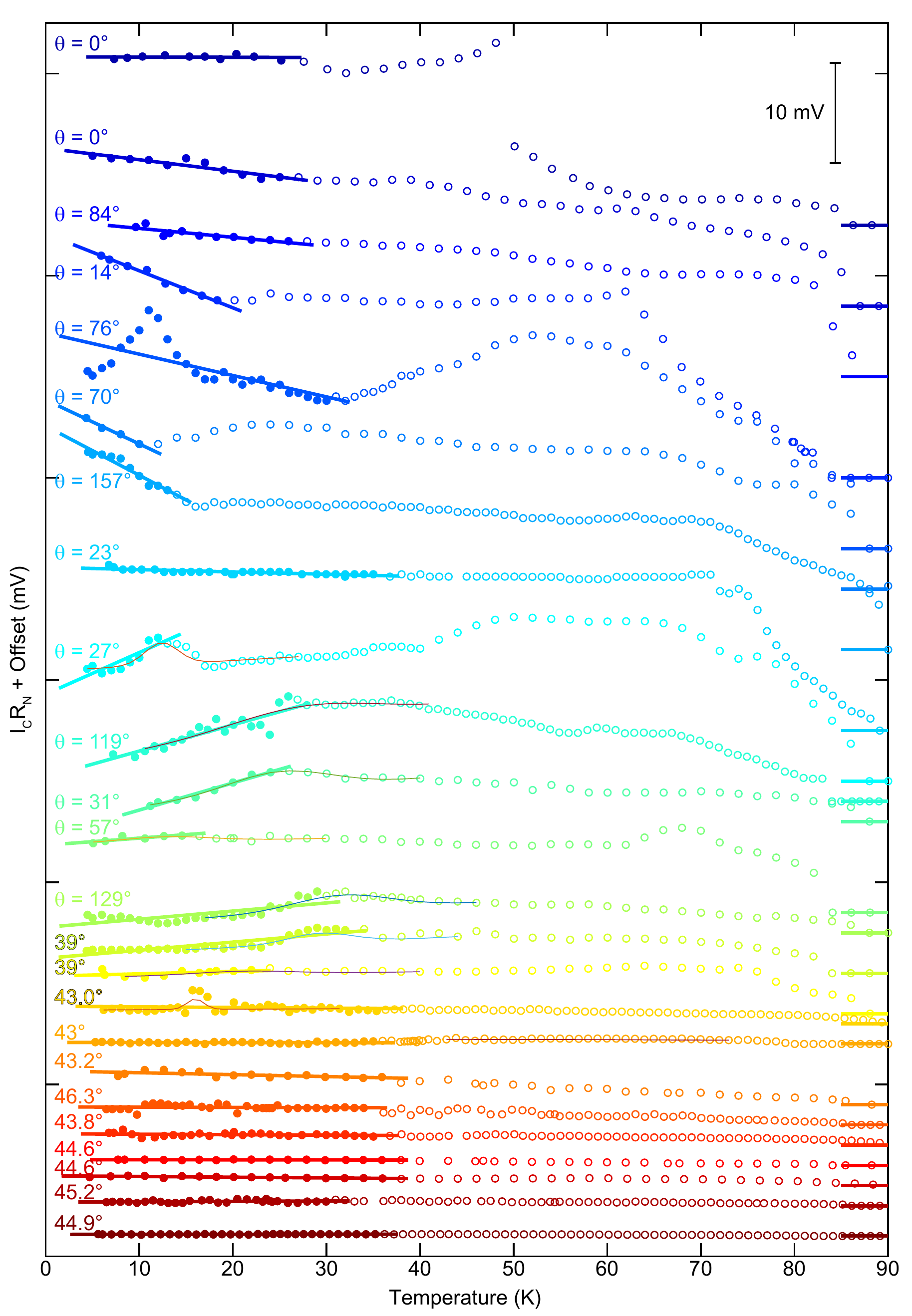}
		\setlength{\belowcaptionskip}{-20pt}
		\vspace{-0.4cm}		
		\caption
		{
			\textbf{$I_CR_N(T)$ for all samples with $T_C > 79$ K}. Solid points are used to extract $d[I_CR_N]/dT$ by linear fit (lines). Red curves fits to the peak in $I_CR_N(T)$. Horizontal lines at right side of plot shows the offset for each curve, where $I_CR_N = 0$. Curves are ordered by $|\tilde\theta - \pi/4|$ where $\tilde\theta = \theta$ mod $\pi/2$. 
		}
		\label{S4}
	\end{center}
\end{figure*}

\section{Theoretical analysis of $I_c(T)$}
Here we analyze the temperature and twist-angle dependence of the critical current in twisted $d$-wave Josephson junctions. We consider two layers of a $d$-wave superconductor weakly coupled by tunneling (as in BSCCO the c-axis anisotropy is very high and the current mostly flows along the interface, we ignore the effects of other layers) described by the Hamiltonian
\begin{equation}\label{es1}
\begin{gathered}
\cH=\sum_{\bf k\alpha}
\xi_{\bk\alpha}\sum_\sigma c^\dagger_{\bk\sigma\alpha} c_{\bk\sigma\alpha}
+
\sum_{\bk\sigma}t({\bk_1 - \bk_2})(c^\dagger_{\bk\sigma1}c_{\bk\sigma2}+c^\dagger_{\bk\sigma 2}c_{\bk\sigma1})
+ \\
+\sum_{\bk}
(\Delta_{\bk 1} e^{i\varphi} c^\dagger_{\bk\uparrow 1}c^\dagger_{-\bk\downarrow 1}+ \Delta_{\bk 2} c^\dagger_{\bk\uparrow 2}c^\dagger_{-\bk\downarrow 2}+ 
{\rm h.c.})
\end{gathered}
\end{equation}
Here $\xi_{\bk \alpha}$ and $\Delta_{\bk \alpha}$ represent the single-particle dispersion and the gap function, respectively, in layer $\alpha=1,2$, related by rotation through twist angle $\theta$, and $\varphi$ is the phase difference between the layers, assumed to be held constant. The twist enters the gap functions, which we assume to depend only on the polar angle in momentum space, via $\Delta_{\bk 1,2}=\Delta\cos(2\alpha_\bk\pm\theta)$, where $\alpha_\bk$ denotes the polar angle of vector $\bk$. The interlayer coupling term allows for momentum non-conserving tunnelling processes, as discussed in more detail below.

\subsection{Nearly incoherent tunneling}
Motivated by the observation of $I_c(\theta)\sim|\cos 2 \theta|$ behavior at low temperatures, characteristic of strongly incoherent tunneling \cite{TwistTheoryRev}, we deduce here the temperature dependence of $I_c(\theta,T)$.
The self-consistent gap equation is given by
\begin{equation}
\frac{1}{\lambda}\equiv
T_c\sum_{\varepsilon_n,{\bf k}} \frac{\cos^2 2\alpha_\bk}{\varepsilon_n^2+\xi_{\bf k}^2}
= T\sum_{\varepsilon_n,{\bf k}} \frac{\cos^2 2\alpha_\bk}{\varepsilon_n^2+\xi_{\bf k}^2+\Delta^2(T) \cos^2 2\alpha_\bk},
\label{eq:selfcon}
\end{equation}
where $\lambda$ represents pairing interaction strength and $\epsilon_n=\pi T(2n+1)$ are fermionic Matsubara frequencies.
It can be directly shown that at low temperatures $\Delta(T)-\Delta(0)\sim T^3$. For a general tunneling amplitude $t({\bf k-k'})$, the interlayer current (in lowest order in $t$) is given by:
\begin{equation}
\begin{gathered}
I(\theta,\varphi) =
T \sum_{\varepsilon_n} \sum_{{\bf k},{\bf k}'} \frac{4 e |t({\bf k-k'})|^2 \Delta^2(T) \sin\varphi \cos 2\alpha_\bk\cos 2(\alpha_{\bk'}+\theta)/\hbar}
{[\varepsilon_n^2+\xi_{\bk1}^2+\Delta^2(T)\cos^2 2\alpha_\bk]
	[\varepsilon_n^2+\xi_{\bk'2}^2+\Delta^2(T)\cos^2 2(\alpha_{\bk'}+\theta)]}.
\end{gathered}
\label{eq:I0q=0}
\end{equation}
For strongly incoherent tunneling we can assume that $\xi_{\bk1}$ and $\xi_{\bk2}$ are independent, while for the angles $\alpha_\bk,\;\alpha_{\bk'}$, we include a finite spread $2\tilde{\sigma}$ with tunneling (i.e.\ $\alpha_\bk$ can tunnel to $[\alpha_\bk-\tilde{\sigma},\alpha_\bk+\tilde{\sigma}]$). Expanding the multiplied factors in \eqref{eq:I0q=0} in Fourier series in $\cos 2 n \alpha_\bk$ and $\cos 2 n (\alpha_{\bk'}+\theta)$, respectively, we find the contributions of higher harmonics to be suppressed due to averaging over the angle. In the limiting case, only the lowest harmonic remains, leading to the final expression
\begin{equation}
\begin{gathered}
I_c^{incoh}(\theta,T) =
\frac{\pi |\cos 2 \theta|\sin 2 \tilde{\sigma} }{2\tilde{\sigma}}
\frac{4 e\nu^2 \overline{t}^2 } {\hbar}
g(T),
\\
g(T) = \frac{4}{\Delta^2(T)} T \sum_{\varepsilon_n}  
\left[\int d \xi 
\left(1- \sqrt{\frac{\varepsilon^2+\xi^2}{\varepsilon^2+\xi^2+\Delta^2(T)}}\right)\right]^2 
\end{gathered}
\label{eq:ictdirty}
\end{equation}
where $\overline{t}^2$ is the average of the tunneling matrix element squared and for $\Delta(T)$ we use the result of a numerical solution of Eq.~\eqref{eq:selfcon}. Note that the $|\cos 2\theta|$ dependence of $I_c(\theta,T)$ is naturally recovered this way. However, the resulting temperature dependence shown in Fig. 2 C of the main text is monotonic. Below we analyze the effects of coherent tunneling that can explain the nonmonotonic $I_c(T)$ behavior.

\subsection{Anomalous temperature dependence of $I_c(T)$: coherent tunneling}

Here we show that a simple model of a $d$-wave superconductor based on Hamiltonian \eqref{es1} generically gives $I_c(T)$ that is non-monotonic and qualitatively similar to experimental data when the interlayer tunneling is momentum-conserving to a good approximation.  To enable analytic progress we assume a simple rotation-invariant dispersion relation $\xi_\bk=\hbar^2 k^2/2m -\mu$ common to both layers and we focus on the case when interlayer tunneling conserves the in-plane momentum and is independent of it. To study the interlayer current it is useful to rewrite the Hamiltonian Eq.\ \eqref{es1}
as $\cH=\sum_\bk\Psi_\bk^\dag H_\bk \Psi_\bk+E_0$
where $\Psi_\bk=(c_{\bk\uparrow  1}, c^\dag_{-\bk\downarrow 1}, c_{\bk\uparrow 2},c^\dag_{-\bk\downarrow 2})^{T}$ represents a four-component Nambu spinor, $E_0$ is a constant and the Bogoliubov-de Gennes (BdG) Hamiltonian is given by a $4\times 4$ matrix
\begin{equation}\label{h4}
  H_\bk=
 \begin{pmatrix}
   \xi_{\bk 1} & \Delta_{\bk 1}e^{i\varphi} & t & 0 \\
   \Delta_{\bk 1}e^{-i\varphi} & -\xi_{\bk 1} & 0 & -t \\
   t & 0 &   \xi_{\bk 2} & \Delta_{\bk 2} \\
   0 & -t &    \Delta_{\bk 2} & -\xi_{\bk 2}
  \end{pmatrix}.
\end{equation}
The interlayer supercurrent can now be obtained from the Josephson relation 
\begin{equation}\label{e5}
  I(\varphi)=(2e/\hbar)d\cF_{\rm BdG}/d\varphi,
\end{equation}
where the free energy $ \cF_{\rm BdG}$ of the system is given by
\begin{equation}\label{h8}
 \cF_{\rm BdG}=E_0-2 k_BT\sum_{\bk a}\ln\left[2\cosh{(E_{\bk a}/2 k_BT)}\right].
\end{equation}
The sum extends over all positive energy eigenvalues $E_{\bk a}$ of the BdG Hamiltonian Eq.\ \eqref{h4} 
which are given by 
\begin{equation}\label{a1}
E_{\bk\pm}=\sqrt{(\Delta_{\bk 1}^2+\Delta_{\bk 2}^2)/2 + \xi_\bk^2 + t^2 \pm D_\bk(\varphi)}
\end{equation}
and $D_\bk^2(\varphi)=(\Delta_{\bk 1}^2-\Delta_{\bk 2}^2)^2/4 +t^2(\Delta_{\bk 1}^2+\Delta_{\bk 2}^2+4\xi_\bk^2-2\Delta_{\bk 1}\Delta_{\bk 2}\cos{\varphi})$. Noting that the phase $\varphi$ only enters through the cosine term in $D_\bk^2(\varphi)$ it is possible, with use of Eq.\ \eqref{e5}, to express the supercurrent as
\begin{equation}\label{a2}
I(\varphi)=-\sin{\varphi}{et^2\over 2\hbar}\sum_\bk{\Delta_{\bk 1}\Delta_{\bk 2}\over D_\bk(\varphi)}\sum_{a=\pm} {a\over E_{\bk a}}\tanh{{1\over 2}\beta E_{\bk a}}.
\end{equation}
This relation is non-perturbative in $t$; setting $t\to 0$ inside the sum one recovers the usual leading-order expression  which is valid to second order in $t$. Unlike the perturbative result discussed above in S5.1, Eq.\ \eqref{a2} gives a small but non-vanishing critical current even at $\theta=45^{\rm o}$, in agreement with experimental observations. It also correctly captures the gapped behavior of the system that occurs when the time-reversal symmetry is broken either spontaneously near $\theta=45^{\rm o}$ or due to externally imposed phase bias. However, the basic phenomenology of the temperature dependence discussed below is contained already in the leading perturbative expression.

To determine $I_c$ it is necessary to find the maximum of $I(\varphi)$ given by Eq.\ \eqref{a2}. Because the maximum is attained at a generic value of $\varphi$ this can generally only be done numerically. We find, however, that for twist angles not too close to 45$^{\rm o}$ the maximum occurs near $\varphi=\pi/2$ and one can approximate $I_c\approx I(\pi/2)$ to a good accuracy. The temperature dependence of the critical current can therefore be usefully analyzed from the expression 
\begin{equation}\label{a3}
I_c(T)\simeq{et^2\over 2\hbar}\sum_\bk{\Delta_{\bk 1}\Delta_{\bk 2}\over D_\bk(\pi/2)}\sum_{a=\pm} \left[{-a\over E_{\bk a}}\tanh{{1\over 2}\beta E_{\bk a}}\right]_{\varphi\to\pi/2}.
\end{equation}
Noting that by definition $E_{\bk +}>E_{\bk -}$ it is easy to show that the last term $\sum_{a=\pm}[\dots]$ in the above equation is non-negative for all temperatures $T$, as is $D_\bk(\pi/2)$. The sign of the contribution of each momentum $\bk$ to the critical current is therefore solely determined by the product of the two $d$-wave gap functions $\Delta_{\bk 1}\Delta_{\bk 2}=\Delta^2\cos(2\alpha_\bk+\theta)\cos(2\alpha_\bk -\theta)$. It is easy to see that for non-zero twist this product is {\em negative} in the vicinity of the Brillouin zone diagonals, i.e.\ the nodal region of the original untwisted $d$-wave superconductor, and is positive in the rest of the BZ. This structure provides for a simple intuitive understanding of the observed decrease in $I_c(0)$ with an increasing twist angle. When $\theta=0$ there are only positive contributions to $I_c(0)$ from the $\bk$ sum and all momenta contribute coherently. On the other hand for $\theta>0$ nodal regions begin to contribute negatively, reducing the critical current and eventually driving it to near zero when $\theta\simeq 45^{\rm o}$. 

The sign structure in Eq.\ \eqref{a3} also helps to explain the anomalous increase in $I_c(T)$ at low temperatures observed for non-zero twist angles. Nonzero temperature promotes existence of pair-breaking excitations which tend to suppress the supercurrent. In a $d$-wave superconductor low-energy excitations reside in the nodal region of the BZ meaning that at low temperatures Cooper pairs composed of electrons with momenta in the nodal region are broken with the highest probability. We argued above, however, that in a twisted configuration nodal regions give a {\em negative} contribution to $I_c(0)$. Reducing this negative contribution by thermal excitations therefore produces a {\em net increase} in the total supercurrent. A detailed analysis of Eq.\ \eqref{a3} given in Ref.\ \cite{Plugge2021} indeed shows an exponentially activated increase in $I_c(T)$ at the lowest temperatures that can be attributed  to a spectral gap $\sim t^2/\Delta$ induced by the $\pi/2$ interlayer phase difference \cite{Volkov}. At temperatures above this small energy scale one finds $I_c(T)\simeq I_c(0)+a_\theta T-b_\theta T^3$ with $a_\theta$ and $b_\theta$ non-negative, twist-angle dependent coefficients. The theoretical analysis thus predicts an approximately linear increase in $I_c(T)$ up to a maximum at $T_{M}=\sqrt{a_\theta/3 b_\theta}$, followed by a decrease at higher temperatures. Coefficient $a_\theta$ is found to grow with increasing $\theta$, reflecting the increasing range of momenta where $\Delta_{\bk 1}\Delta_{\bk 2}<0$.  This causes the position of the maximum $T_{M}$ to shift to higher temperatures for larger $\theta$, giving rise to a behavior that is qualitatively consistent with $I_c(T)$ measured in our twisted junctions.   

\section{Voltage Jump at $I_C$}
The $I$-$V$ characteristics features a jump at $I_C$, which provides an independent way to measure the symmetry of the superconducting order parameter. Figure S5 shows the twist angle dependence of $V(I_C)$ for all devices, at different temperatures. $V(I_C)$ closely matches the expected $|\cos (2\tilde \theta)|$ dependence expected of SIS Josephson junctions between $d$-wave superconductors.

\begin{figure*}[h!]
	\vspace{-0.5cm}
	\begin{center}
		\includegraphics[width=0.7\linewidth]{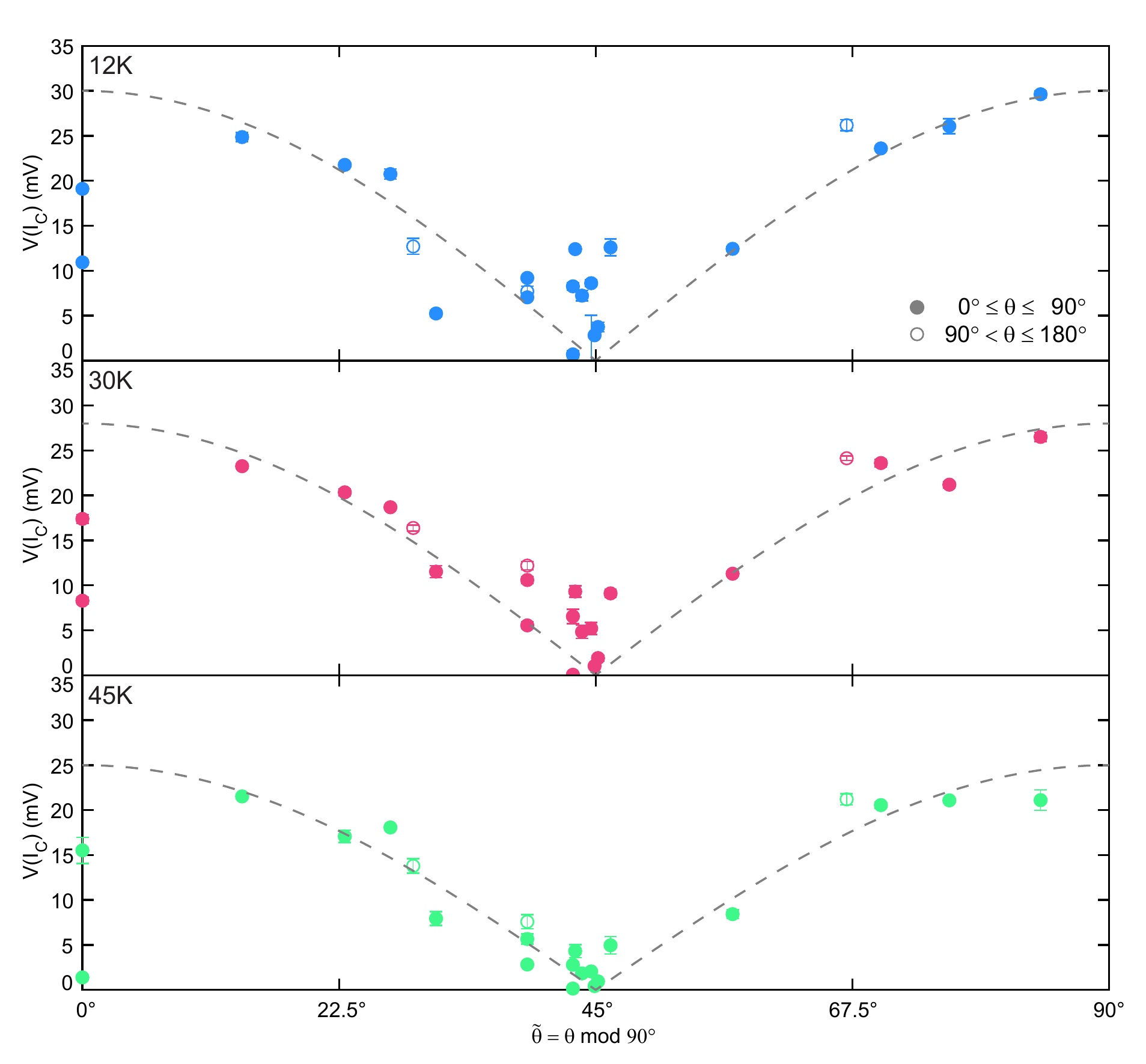}
		\setlength{\belowcaptionskip}{-20pt}
		\vspace{-0.4cm}		
		\caption
		{
			\textbf{$V(I_C)$ vs $\tilde \theta$} at 12, 30 and 45~K. Dashed line shows the $|\cos (2\tilde \theta)|$ dependence expected for JJs between $d$-wave superconductors.
		}
		\label{S5}
	\end{center}
\end{figure*}

\section{Shapiro Step Frequency Dependence}
\begin{figure*}[h!]
	\vspace{-0.5cm}
	\begin{center}
		\includegraphics[width=0.7\linewidth]{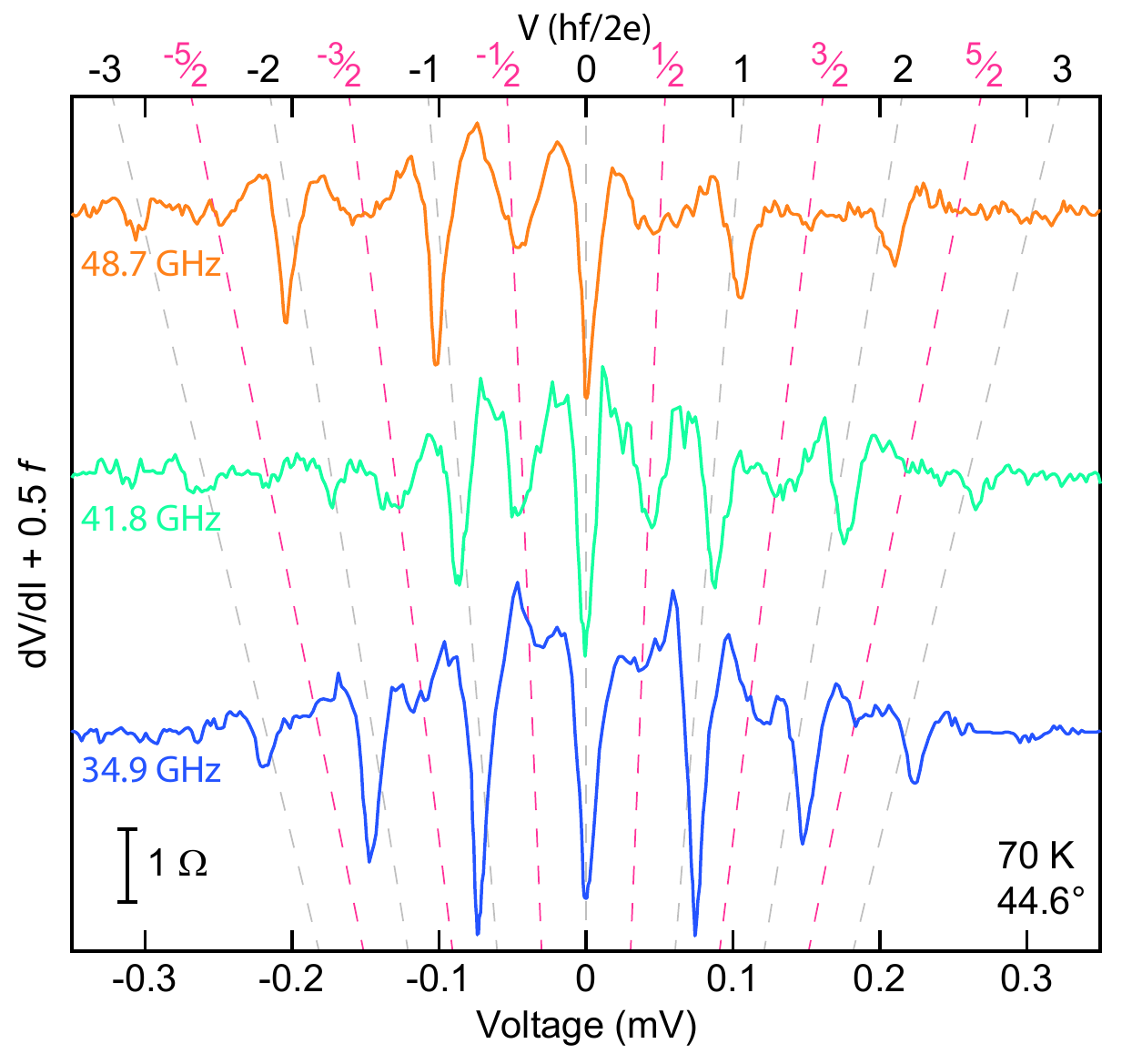}
		\setlength{\belowcaptionskip}{-20pt}
		\vspace{-0.4cm}		
		\caption
		{
			\textbf{Shapiro step frequency dependence}, showing junction $dV/dI$ vs. $V$ under microwave illumination at frequency $f$. Each trace is offset by $0.5f$, so that the expected position of each Shapiro step follows the dashed lines. Half-integer steps are highlighted in red. 
		}
		\label{ShapiroFreq}
	\end{center}
\end{figure*}

Shapiro steps are expected to appear at multiples $n$ of $hf/2e$, where $f$ is the illuminating microwave frequency. We explicitly check this dependence in Figure \ref{ShapiroFreq}, where the periodic $dV/dI$ dips fan out linearly with $f$, at both integer and half-integer $n$, as expected. This is evidence that the observed $dV/dI$ dips are indeed Shapiro steps. 

\section{Exclusion of Alternative Mechanisms of Half-Integer Shapiro Steps}
The magnetic fields enclosed in Josephson junctions are expected to vary on scale of the Josephson length $\lambda_J$ \cite{Barone}. Junctions with lateral size exceeding $\lambda_J$ may contain Josephson vortices, whose dynamics can also induce subharmonic Shapiro steps~\cite{terpstra1995} at fractional multiples of $hf/2e$. Such steps usually appear at nonzero magnetic fields in large junctions. 

We can estimate the Josephson length by the expression $\lambda_J = \sqrt{\hbar / 2 e \mu_0 J_C d}$, where $J_C$ is the critical current density and $d$ is the thickness of the bulk crystal surface layer where magnetic fields may penetrate \cite{Barone}. From our Fraunhofer patterns, $d \approx 30$ nm and is bounded above by the total thickness of the device. In our 44.6$^\circ$ junction at 70~K where the data for Figure 4 is taken, the critical current density is 9.7 A / cm$^2$, which corresponds to $\lambda_J = 300~\mu$m. This is 10 times larger than the actual lateral size of the device, which is about 25 $\mu$m on the longest axis. For such small devices, subharmonic Shapiro steps originating from flux dynamics are not expected to appear. 

In addition, our Shapiro steps are detected at zero magnetic field, and only appear when the twist angle $\theta$ is within about a 1$^\circ$ range around $45^\circ$. This is precisely the angle where $J_C$ is maximally suppressed and $\lambda_J$ reaches a maximum. We therefore conclude that the half-integer Shapiro steps are unlikely to originate from flux motion.

At the temperature where half-integer Shapiro steps are presented in Figure 4E at $44.6^\circ$ and $45.0^\circ$ (at 70 K and 65 K respectively), we observe no hysteresis in the $I$-$V$ curve. 

\section{Fraunhofer patterns near 45$^\circ$}
\subsection{Qualitative assessment}

We model the current density-phase relation near 45$^\circ$ with two sinusoidal harmonics:
\begin{equation}
j_c(\theta,\varphi) = j_c^1(\theta) \sin(\varphi(x)) - j_c^2 \sin(2\varphi(x)),
\label{eq:jc}
\end{equation}
where the first term describes the Cooper pair tunneling, required to vanish at 45$^\circ$, while the second one describes a higher-order process: co-tunneling of Cooper pairs. We assume $j_c^2\ll j_c^1(\theta=0)$ due to the smallness of the interlayer tunneling at the interface. In a magnetic field the phase becomes position-dependent, with its characteristic variation length in the junction being given by the Josephson length $\lambda_J\sim 1/\sqrt{j_c}$ \cite{Barone}. To discuss the case above we introduce two Josephson lengths $\lambda_{J1}(\theta)\sim 1/\sqrt{|j_c^1(\theta)|}$, $\lambda_{J2}\sim 1/\sqrt{j_c^2}$.

\begin{figure}[h]
	\begin{center}
		\includegraphics[width=0.8\textwidth]{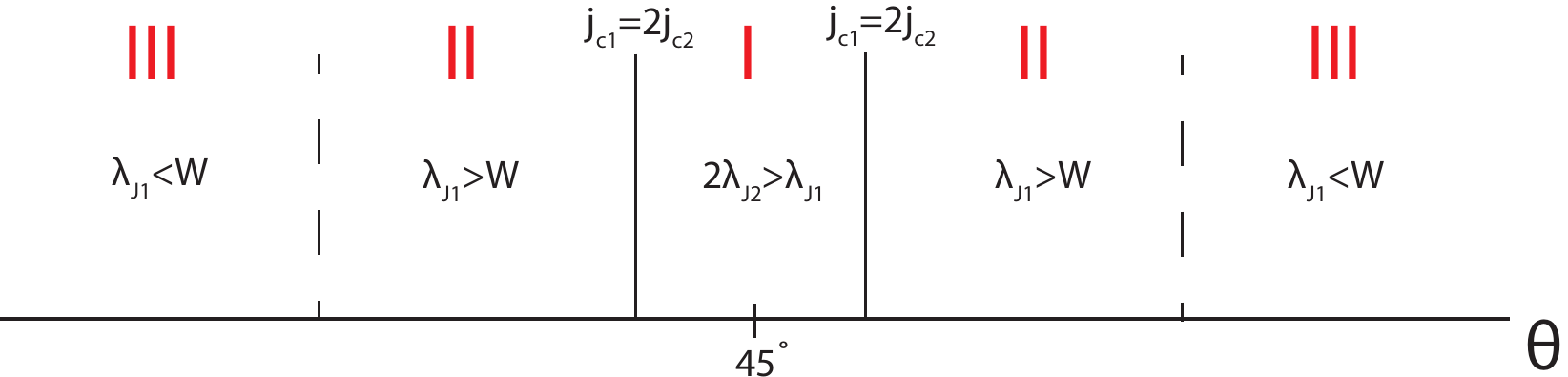}
	\end{center}
	\caption{Regimes of a $d$-wave twist junction near 45$^\circ$; only in regimes $II$ and $I$ a Fraunhofer pattern is expected. Line separating $II$ and $I$ corresponds to phase transition to a topological phase in regime $I$.
	}
	\label{fig:supregimes}
\end{figure}

Close to 45$^\circ$, three regimes are possible (Fig. \ref{fig:supregimes}), depending on the relation between $\lambda_{J1,2}$ and the junction overlap length $w$.  Due to $j_c^2\ll j_c^1(0)$, we assume that $\lambda_{J2}\gg w$. Furthest away from 45$^\circ$ is regime $III$: there, the Josephson length $\lambda_{J1}(\theta)$ becomes shorter than $w$. In this case, the system is not expected to exhibit a Fraunhofer interference pattern (FIP) in a magnetic field \cite{Barone}, showing instead a monotonic decrease of $I_c$ with increasing field. The FIP first appears in regime $II$, where the first harmonic in CPR is dominant. The line $j_c^1(\theta) = 2 j_c^2$ corresponds to a topological phase transition. The free energy density $U(\theta,\varphi) \sim \int_0^\varphi d\varphi' j_c(\theta,\varphi)$ has a minimum at $\varphi=0$ in regime $II$, but in regime $I$, the minimum shifts to a non-zero value. The non-zero phase difference breaks the time-reversal symmetry of the ground state and results in topological superconductivity \cite{canHightemperatureTopologicalSuperconductivity2021}.

\subsection{Model for magnetic field effects}

\begin{figure}[h]
	\begin{center}
		\includegraphics[width=0.8\textwidth]{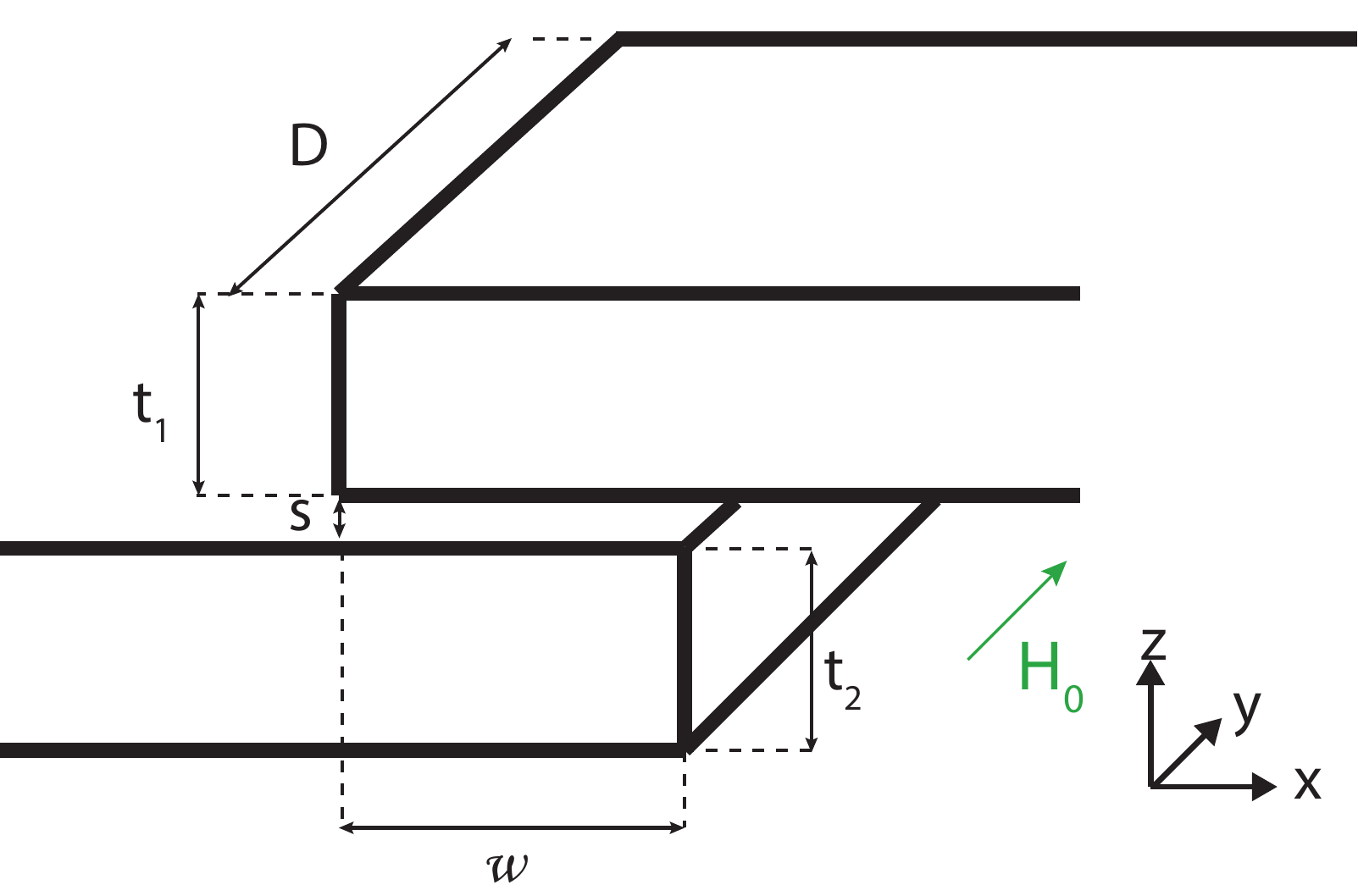}
	\end{center}
	\caption{Junction geometry considered; the magnetic field $H_0$ is along $y$ axis and $D\gg W$ is assumed. 
	}
	\label{fig:geom}
\end{figure}

We consider the model geometry presented in Fig. \ref{fig:geom} with field along $y$. Assuming the depth of JJ $D$ to be much larger than the width $w$, we ignore the $y$-dependence of the field, reducing the problem to a two-dimensional one for $x$ and $z$ only. The FIP in Fig. 3 B,C of the main text suggests that the JJs are in regime $I$ or $II$, allowing us to neglect the self-field effects of the junction. In this case, the phase difference at the twist junction $\varphi(x)$ in a magnetic field $H(x,z)$ satisfies:

\begin{equation}
\frac{\partial \varphi}{\partial x}(x)
=
\frac{2\pi s}{\Phi_0} H_0+
2\pi \lambda_{ab}^2\left[
\left.\frac{\partial H (x,z)}{\partial z}\right|_{z=z_2^{top}}
-
\left.\frac{\partial H(x,z)}{\partial z}\right|_{z=z_1^{bot}}
\right],
\label{eq:phix}
\end{equation}
where $H_0$ is the applied external field $z_2^{top}=t_2$ corresponds to the top surface of the lower flake and $z_1^{bot}=t_2+s$ to the bottom surface of the upper flake.

Secondly, we assume the external magnetic field to be small enough such that the layered nature of the BSCCO flakes and the presence of vortices therein can be ignored as the characteristic scales for an intrinsic junctions are of the order $0.1$ T \cite{latyshevDimensionalCrossoverIntrinsic1996}. As for the vortices, for magnetic field strengths $H\lesssim H_{vort}\sim \frac{\Phi_0}{W t_{1,2}}$, less than one vortex is close to the junction area. For typical $w\sim$10~$\mu$m and $d\sim$0.05-0.1~$\mu$m, $H_{vort}$ is between $20$ and $40$ Gauss, allowing to neglect their presence for order of magnitude estimates, as only very few of them are present in the vicinity of the junctions at the relevant fields.

Under the above conditions, the magnetic field $H(x,z)$ inside a rectangular flake of thickness $d$ and length along $x$ coordinate $L$ satisfies London equations:
\begin{equation}
\begin{gathered}
\lambda_c^2 \frac{\partial^2 H}{\partial x^2}
+
\lambda_{ab}^2 \frac{\partial^2 H}{\partial z^2}
=
H,
\\
H|_{z=\pm d/2} = H_0; H|_{x=\pm L/2} = H_0,
\end{gathered}
\label{eq:londeq}
\end{equation}
where $H(x,z)$ is the magnetic field within the flake, $\lambda_{ab}$, $\lambda_c$ are the London penetration depths in the $a-b$ plane and along the $c$-axis, respectively, $H_0$ is the applied external field value.
Solving the equation above for each flake in Fig. \ref{fig:geom} (note that the flakes are shifted along $x$ in the overlap junction configuration) one can obtain $\varphi(x)$ by integrating Eq.~\eqref{eq:phix}. However, for order of magnitude estimate it is convenient to use the average value of $H(x,z)$ over $x\in[0,w]$ rather then the full $x$-dependent function. The averaging is a good approximation when $\frac{\pi w \lambda_{ab}}{d \lambda_c}\ll1$ (which is justified for our experimental system, see below). Furthermore, we assume that $\frac{\pi\lambda_{ab}}{d}$ can be taken to be much larger than $1$ as the flake's thicknesses are below 100 nm, while $\lambda_{ab}\sim 0.2\mu$m \cite{enriquez2001} and 
we assumed the length of the flakes along the $x$ coordinate $L$ to be much larger than $t_{1,2} \lambda_c/(\pi\lambda_{ab})$.
The dependence $\varphi(x)$ is then given by:
\begin{equation}
\varphi(x)\approx \frac{2 \pi H_0 d x}{\Phi_0}+C,
\end{equation}
where
\begin{equation}
\begin{gathered}
d
=
s
+
\sum_{i=1,2}
\left(
\lambda_{ab}\tanh \frac{t_i}{2 \lambda_{ab}} 
-
\frac{4 t_i}{\pi^2}
\sum_{n=0}^\infty
\frac{1}{(2n+1)^3}
\frac{1-
\exp\left(-(2n+1)\frac{w\pi\lambda_{ab}}{d_i\lambda_c}\right)
}
{\frac{w\pi\lambda_{ab}}{d_i\lambda_c}}
\right),
\end{gathered}
\label{eq:approx}
\end{equation}
which depends on two dimensionless parameters: $\frac{t_i}{2 \lambda_{ab}}$ and $\frac{w\pi\lambda_{ab}}{t_i\lambda_c}$. For a purely first-harmonic dominated current-phase relation (i.e. $j_c^2=0$ in Eq.~\eqref{eq:jc}) one obtains then the conventional Fraunhofer pattern, with the first zero being at a field:
\begin{equation}
H^{(1)}_0 = \frac{\Phi_0}{w d},
\label{eq:expdeff1}
\end{equation} 
which allows us to extract the value of $d$ from the experimentally observed Fraunhofer pattern. 
Note that in the opposite case $j_c^1=0$ (i.e. at $\tilde{\theta}=45^\circ$) the first zero in the pattern occurs at 
\begin{equation}
H^{(2)}_0 = \frac{\Phi_0}{2 w d},
\label{eq:expdeff2}
\end{equation}
which implies a twice smaller $d$ value for the same Fraunhofer pattern. When both $j_c^1$ and $j_c^2$ are nonzero, the dependence $I_c(H)$ interpolates between the two limits, with the odd-numbered zeros of the $j_c^2$-dominated FIP being gradually lifted.

\begin{figure}[h!]
	\begin{center}
		\includegraphics[width=0.6\textwidth]{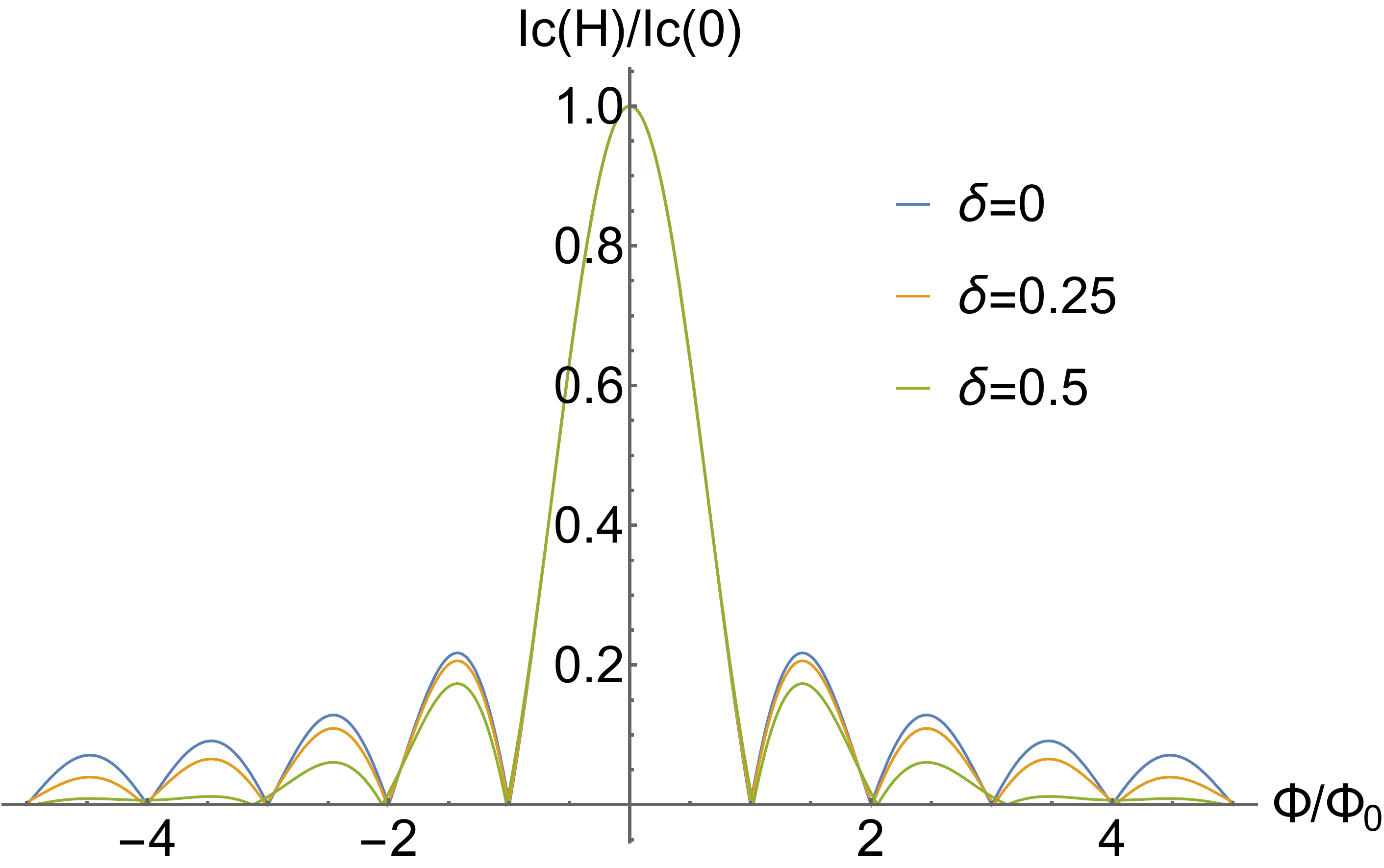}
	\end{center}
	\caption{Effect of the variation of the effective thickness on the Fraunhofer pattern. A box distribution for $d$ is assumed $d_{\mathrm{eff}} \in d_{\mathrm{eff}}^0[1-\delta/2,1+\delta/2] $.}
	\label{fig:Fr3}
\end{figure}

For non-rectangular junctions, $w$ varies along the junction depth ($y$ coordinate), leading, in turn, to variation of $d(y)$. The former effect can lead to a deformation of the Fraunhofer pattern, whereby zeroes will be not at the positions expected from rectangular geometry \cite{Barone}, while the second one leads to a suppression of the $I_c$ at large field value due to destructive interference along the depth of the junction (see Fig. \ref{fig:Fr3}). The latter observation is consistent with experiments.

\subsection{Comparison with experiment}
We now apply the findings above to the experimental results.
%
The simplest case to analyze is the short-period pattern in Fig. 3B of the main text (we discuss the second, "slow" feature below). $I_cR_N$ is an order of magnitude larger here than that at $44.9^\circ$ (Fig. 3C of the main text). As only $j_c^1$ has a strong angular dependence close to $45^\circ$ the Fraunhofer pattern likely corresponds to the case $j_c^1\gg j_c^2$. The value of $d$ resulting from Eq.~\eqref{eq:expdeff1} is $15$~nm. Using the AFM measured actual thicknesses of the flakes $t_1=t_2=80$~nm and $w=10.4\;\mu$m we deduce $\frac{\lambda_c}{\lambda_{ab}}(T=20\;K)\approx1.8 \cdot 10^3$ from Eq.~\eqref{eq:approx}, somewhat larger than in single crystal whiskers \cite{latyshevDimensionalCrossoverIntrinsic1996}. While this number may not reflect the actual penetration depth anisotropy due to the simplified model of geometry we consider, it represents an intrinsic characteristic of the flakes, and as such should not be dependent on twist angle. On the other hand, the temperature-dependence of $d$ can be understood (from \eqref{eq:approx}) to originate from the temperature dependence of the anisotropy $\frac{\lambda_c}{\lambda_{ab}}(T)$. Indeed, the penetration depths $\lambda_c(T)$ and $\lambda_{ab}(T)$ show different dependence on the temperature $T$ \cite{enriquez2001}.

The coexistence of two critical-current like features strongly resembles the situation in systems of two junctions in series \cite{nevirkovets1994}. In that case, a short ($\lambda_J\gg w$) junction is in series with a long junction ($\lambda_J\ll w$). As the critical current of the first one is strongly reduced by field, the change in the geometry of current flow also reduces the critical current of the second one, which produces a slower decreasing critical-current-like feature.

At $44.9^\circ$ (Fig. 3 C of the main text) a rather clear Fraunhofer pattern is observed, implying the dominance of either $j_c^1$ or $j_c^2$.
In the following we consider both cases
$j_c^1\gg j_c^2$ and $j_c^2\gg j_c^1$ to determine which is most consistent with the experimental data.

For $j_c^2\gg j_c^1$, we find that
$d\approx 12$ nm from Eq.~\eqref{eq:expdeff2} 
at $T=20$ K (Fig. 3 C of the main text, right inset) resulting in an estimate $\frac{\lambda_c}{\lambda_{ab}}(T=20\;K)\approx 2.3 \cdot 10^3$, roughly consistent with the value deduced from Fig. 3 B of the main text $(\approx 1.8 \cdot 10^3)$ with $d_1=d_2=66$ nm deduced from AFM measurements. It reduces at higher temperature to $\frac{\lambda_c}{\lambda_{ab}}(T=60\;K)\approx 1.2 \cdot 10^3$. Importantly, close to $T_c$, one expects $j_c^1(T)\sim |\Delta|^2\sim(T-T_c)$, while $j_c^2(T)\sim |\Delta|^4\sim(T-T_c)^2$ and hence a crossover to the   regime $j_c^2\ll j_c^1$ 
at high temperatures is possible that
would result in lifting of odd-numbered zeros. However, such a behavior is not observed,
which is consistent with the fact that
 the $\pm 0.1^\circ$ uncertainty of the twist angle leaves room for arbitrary small values of $j_c^1$, limiting the crossover temperature to an unobservably small vicinity of $T_c$.

If we instead assume that $j_c^2\ll j_c^1$ for the $44.9^\circ$ junction, the value of $d_{\mathrm{eff}}$ from Eq.~\eqref{eq:expdeff1} is around $24$ nm at $T=20$ K, resulting in an estimate $\frac{\lambda_c}{\lambda_{ab}}(T=20\;K)\approx 0.8 \cdot 10^3$, which is further from the value deduced from Fig. 3 B of the main text, than the one deduced assuming  $j_c^2\gg j_c^1$. Note that even if this case is realized, a substantial $j_c^2\sim j_c^1$ is consistent with the observations, as the FIP for $2j_c^2=j_c^1$ is almost indistinguishable from the one at $j_c^2\ll j_c^1$.

Finally, Fig. 3 A does not show a clear Fraunhofer-like pattern and is furthest from $45^\circ$. Its critical current density is smaller than that of Fig. 3 B, which suggests the presence of disorder that relaxes the in-plane momentum conservation at the interface. At the same time the junction width $w$ is smaller than in Fig. 3 B, implying that a crossover to a long-junction limit (where the $\lambda_J\sim1/\sqrt{j_c}$ is smaller than $w$) is unlikely. On the other hand, the reduced value of $w$ leads to a larger period of the Fraunhofer pattern. Using the lowest anisotropy value from the ones deduced above at $T=45$~K, $\frac{\lambda_c}{\lambda_{ab}}(T=40$~K$)> 0.6 \cdot 10^3$ we deduce $d<17$ nm ($t_1=64$ nm, $t_2=124$ nm deduced from AFM measurements) and $H^{(1)}_0 > 300$ Gauss, much larger than in the other samples. A possible scenario is then that the Fraunhofer pattern is smeared by the field inhomogeneities created by vortices in the flakes \cite{fistul1994}, that can not be neglected at such high fields (note that the characteristic field, where suppression becomes significant does not depend on junction width $w$ and has been found to be around $200$ Gauss \cite{latyshevDimensionalCrossoverIntrinsic1996}). Indeed, $\frac{\Phi_0}{w t_{1,2}} = 92$ and $47$ Gauss, respectively, suggesting that many vortices will be present in the near-junction region for fields, where the Fraunhofer zero is expected to occur.



\end{document}